\documentclass{article}

\usepackage{PRIMEarxiv}

\usepackage[utf8]{inputenc} % allow utf-8 input
\usepackage[T1]{fontenc}    % use 8-bit T1 fonts
\usepackage{hyperref}       % hyperlinks
\usepackage{url}            % simple URL typesetting
\usepackage{booktabs}       % professional-quality tables
\usepackage{amsfonts}       % blackboard math symbols
\usepackage{nicefrac}       % compact symbols for 1/2, etc.
\usepackage{microtype}      % microtypography
\usepackage{lipsum}
\usepackage{fancyhdr}       % header
\usepackage{graphicx}       % graphics
\graphicspath{{media/}}     % organize your images and other figures under media/ folder
\usepackage{algorithm}
\usepackage{algpseudocode}
\usepackage{amsmath}%
\usepackage{ragged2e}
\usepackage{subcaption}
\usepackage{bm}

\title{Meta-PINN: Meta learning for improved neural network wavefield solutions
}

\author{
  Shijun Cheng and Tariq Alkhalifah \\
  King Abdullah University of Science and Technology, Thuwal 23955-6900, Saudi Arabia. \\
  \texttt{sjcheng.academic@gmail.com, tariq.alkhalifah@kaust.edu.sa} \\
}

\begin{document}
\maketitle

\begin{abstract}
Physics-informed neural networks (PINNs) provide a flexible and effective alternative for estimating seismic wavefield solutions due to their typical mesh-free and unsupervised features. However, their accuracy and training cost restrict their applicability. To address these issues, we propose a novel initialization for PINNs based on meta learning to enhance their performance. In our framework, we first utilize meta learning to train a common network initialization for a distribution of medium parameters (i.e. velocity models). This phase employs a unique training data container, comprising a support set and a query set. We use a dual-loop approach, optimizing network parameters through a bidirectional gradient update from the support set to the query set. Following this, we use the meta-trained PINN model as the initial model for a regular PINN training for a new velocity model in which the optimization of the network is jointly constrained by the physical and regularization losses. Numerical results demonstrate that, compared to the vanilla PINN with random initialization, our method achieves a much fast convergence speed, and also, obtains a significant improvement in the results accuracy. Meanwhile, we showcase that our method can be integrated with existing optimal techniques to further enhance its performance.
\end{abstract}

\section{Introduction}
Seismic modeling plays a crucial role in understanding the Earth's subsurface structure and identifying geological features, and thus, it has broad applications in various fields such as global seismology, near surface geotechnical investigations, and seismic exploration. Seismic modeling is often achieved by solving the wave equation, since it is the fundamental equation that governs the propagation of seismic waves in the subsurface. Accurate and efficient wavefield solutions to the seismic wave equation can help to interpret seismic data and predict the behavior of seismic waves in complex geological structures \cite{aki1980quantitative}. 

Traditional numerical solvers, such as finite element methods (FEMs) and finite difference methods (FDMs), have been widely used for solving seismic wave equations in inhomogeneous media over the years \cite{virieux1986p, de2007grid, virieux2011review, cheng2021wave}. FEMs use a mesh of small elements to approximate the solution by solving the equation at each node, while FDMs approximate the derivatives of the wave equation using the difference between grid points in the domain. They are easy to implement and have good stability properties. However, both methods require the discretization of the seismic wave equations on a fixed grid, which can limit their ability to adapt to new domains and models. Moreover, to achieve an accurate representation of the wavefield solution, a fine-resolution grid is typically necessary. This requirement results in high computational costs, making the numerical simulation impractical for large-scale problems. 

Recently, physics-informed neural networks (PINNs) \cite{raissi2019physics} have gained significant attention for their ability to solve complex partial differential equations (PDEs). The network, like a function, accepts spatial (or spatial-temporal) coordinate values as input and aims to output the value of the wavefield at the input coordinate location. The principle behind PINNs is the intorduction of the governing PDE as a loss function in the neural network (NN)'s training process. By minimizing the loss function, the NN (function) learns the underlying physics of the system and produces accurate predictions of the solution. One of the significant advantages of PINNs is their grid-free nature, which eliminates the need for a predefined grid. This feature allows for easy handling of complex geometries and irregular boundaries while overcoming the discretization error typical of finite difference solvers \cite{taufik2022upwind}. It means that PINNs meshless feature eliminates the need for a fine-resolution grid, which is often required in traditional numerical methods and can lead to considerable computational cost. Another advantage comes from ingenious physical loss, which means that PINNs can work in an unsupervised fashion, avoiding to generate labeled data from conventional supervised learning methods. This also implies that PINNs can possibly adapt to complex equations in which we have failed to solve numerically \cite{taufik2022upwind}.

In the field of seismic modeling, PINNs have shown significant potential due to their ability to simulate wave propagation in complex media \cite{song2021solving, song2022versatile, alkhalifah2021wavefield, huang2022pinnup, karimpouli2020physics, rasht2022physics,brandolin2023pinnslope, taufik2023latentpinns}. The velocity model is a crucial input parameter in solving the seismic wave equation, as it significantly affects the resulting wavefield solution. However, PINN treats each velocity model as a separate modeling task, which renders the previously trained network ineffective for a new velocity model. Consequently, the network needs to be retrained from scratch every time. This process can be computationally expensive and impractical when dealing with a large number of various velocity models, like in iterative inversion applications. To overcome this issue, some preliminary studies suggest using transfer learning algorithms to accelerate the convergence of PINNs. For example, Waheed et al. \cite{bin2021pinneik} proposed initializing PINN models with previously trained network weights, allowing PINNs to converge more quickly on new source positions or velocity models. However, transfer learning presupposes similarity between the source and target domains. Autually, subsurface velocity distributions are complex and varied. New velocity models may differ significantly from those processed by the pretrained network, making it challenging for the pretrained network weights to effectively transfer to modeling tasks involving new velocity models. Furthermore, in order to obtain an ideal wavefield solution, a network architecture with a large capacity is necessary, resulting in a sufficiently large number of parameters that need to be optimized during the training process. If the initialization of these parameters is not appropriate, or if the training process is not carefully designed, the network will spend a considerable amount of time to converge, and also, easily fall into a trivial solution. Additionally, since the loss function used in PINN training is highly nonlinear and non-convex, the optimization process may also fall into the wrong local minimum, which can further exacerbate the problem of trivial solutions. Hence, careful initialization of network parameters is crucial to improve the convergence speed, to avoid trivial solutions, and to ensure that the resulting wavefield solution accurately describe the underlying wave phenomenon.

Meta learning (MetaL) is a powerful technique in machine learning that aims to improve the performance of NN on new tasks and domains efficiently by learning how to learn \cite{finn2017model, hospedales2021meta}. By using the meta-trained model as a starting point, MetaL can reduce the amount of time and compute resources required to achieve state-of-the-art performance on a new task. However, in the field of seismology, limited attention has been devoted to the application of MetaL. For example, Yuan et al. \cite{yuan2020adaptive} tackled the issue of first arrival picking across various seismic datasets by utilizing MetaL algorithms, showcasing results that surpassed those achieved through transfer learning techniques in terms of accuracy. Conversely, Sun and Alkhalifah \cite{sun2020ml} applied the principles of MetaL in formulating an optimization algorithm aimed at enhancing the efficiency of full waveform inversion by significantly speeding up its convergence. Cheng et al. \cite{cheng2023meta} formulated a comprehensive framework for diverse seismic processing tasks, termed Meta-Processing. This approach leverages a minimal amount of training data for meta learning a common network initialization, which offers universal adaptability features.

Drawing inspiration from the MetaL concept, we share an innovative approach to guide PINNs to learn wavefield solutions faster and more accuretly for various velocity models. In this context, we aspire for a PINN solution that represents the general structure of wavefield solutions regardless of the velocity model within a distribution of velocity models. Thus, a plausible assumption is that the function spaces mapped by independent PINNs may have certain underlying connections, for example, a particular operator. Specifically, our objective in this paper is to utilize MetaL to find an initial NN model exhibiting robust generalization ability, which is meta-learned across a limited amount of velocity models, allowing for rapid adaptation in obtaining wavefield soulutions for any arbitrary velocity model. Here, we focus on applying the PINN to solve the wave equation in 2D acoustic isotropic media for scattered wavefields. 

The rest of paper is organized as follows. We first review the wave equation for scattered wavefields, and also, briefly introduce the theoretical concepts behind PINNs. Then, we present the MetaL initialization-based PINN (Meta-PINN) algorithm, as well as the multi-loss constraints method. Following that, numerical examples are shared to validate our method. Furthermore, we discuss some key features, configurations and limitations of our method. Finally, we share our concluding remarks.

\section{Method}
\subsection{Frequency-domain acoustic wave equation}
Seismic waves propagating in an acoustic, isotropic, constant density medium can be simulated by solving the following frequency-domain wave equation:
\begin{equation}\label{eq1}
\omega^2 m u (\bm{\mathrm{x}},\omega,\bm{\mathrm{x_s}} ) +  
\nabla^2 u (\bm{\mathrm{x}},\omega,\bm{\mathrm{x_s}} ) = s(\bm{\mathrm{x}},\bm{\mathrm{x_s}}),
\end{equation}
where $\omega$ is the angular frequency, $m = 1/v^2$ is the squared slowness given by velocity $v$, $u(\bm{\mathrm{x}},\omega,\bm{\mathrm{x_s}})$ is a complex wavefield in the frequency domain, ${\bm{\mathrm{x}}=\left(x,z \right)}$ represents a vector of spatial coordinates for the 2-D medium, $\bm{\mathrm{x_s}}$ denotes the coordinate of the source, ${s(\bm{\mathrm{x}},\bm{\mathrm{x_s}})}$ is the source function, and $\nabla$ represents the Laplacian operator given by the form of $\partial^2/\partial x^2 + \partial^2/\partial z^2$. 

For a point source, equation \ref{eq1} gives rise to singularity in the wavefield solution at the source location, leading to inaccuracies in numerical solutions near the source \cite{alkhalifah2021wavefield}. Moreover, due to the extreme sparsity of the source vector, collecting sufficient samples for training a PINN becomes a daunting task \cite{song2022simulating}. Therefore, as recommended by Song et al. \cite{song2021solving}, we consider scattering theory to represent equation \ref{eq1} in a perturbation form as
\begin{equation}\label{eq2}
\setlength{\abovedisplayskip}{3pt}
\setlength{\belowdisplayskip}{3pt}
\begin{split}
\omega^2 m(\bm{\mathrm{x}}) \delta u (\bm{\mathrm{x}},\omega,\bm{\mathrm{x_s}} ) +  
\nabla^2 \delta u (\bm{\mathrm{x}},\omega,\bm{\mathrm{x_s}} )  = -\omega^2 \delta m(\bm{\mathrm{x}})  u_0 (\bm{\mathrm{x}},\omega,\bm{\mathrm{x_s}} ),
\end{split}
\end{equation}
where $u_0(\bm{\mathrm{x}},\omega,\bm{\mathrm{x_s}})$ represents the background wavefield corresponding to the background velocity $v_0$, $\delta m(\bm{\mathrm{x}}) = m(\bm{\mathrm{x}}) - m_0(\bm{\mathrm{x}})= 1/v^2 - 1/v_0^2$ is the squared slowness perturbation, and $\delta u (\bm{\mathrm{x}},\omega,\bm{\mathrm{x_s}} )$ denotes the scattered wavefield. For the scattered wavefield, the source function in the right-hand side of equation \ref{eq2} directly depends on the perturbation model $\delta m(\bm{\mathrm{x}})$ and background wavefield $u_0(\bm{\mathrm{x}},\omega,\bm{\mathrm{x_s}})$, which often extends the full spatial domain.

Commonly, the background model is considered as an infinite acoustic isotropic medium with a constant velocity. In this case, the background wavefield $u_0(\bm{\mathrm{x}},\omega,\bm{\mathrm{x_s}})$ can be calculated with an analytical formula \cite{aki1980quantitative}:
\begin{equation}\label{eq3}
\setlength{\abovedisplayskip}{3pt}
\setlength{\belowdisplayskip}{3pt}
u_0 (\bm{\mathrm{x}},\omega,\bm{\mathrm{x_s}} )= \frac{\mathrm{i}}{4} H_0^{(2)}\left(\frac{\omega}{v_0}|\bm{\mathrm{x}}-\bm{\mathrm{x_s}}| \right),
\end{equation}
where $H_0^{(2)}$ represents the zero-order Hankel function of the second kind, and $\mathrm{i}$ is the imaginary identity.

\subsection{PINN}
Neural networks have demonstrated their capacity to effectively represent arbitrary nonlinear functions \cite{hornik1989multilayer}. Raissi et al. \cite{raissi2019physics} demonstrated the flexibility of PINNs in representing the solutions of nonlinear PDEs, using the concept of automatic differentiation to evaluate derivatives during backpropagation process. Here, PINN is used to solve equation \ref{eq2} to obtain the desired scattered wavefield solutions. 

\begin{figure}[!t]
\centering
\includegraphics[width=0.98\textwidth]{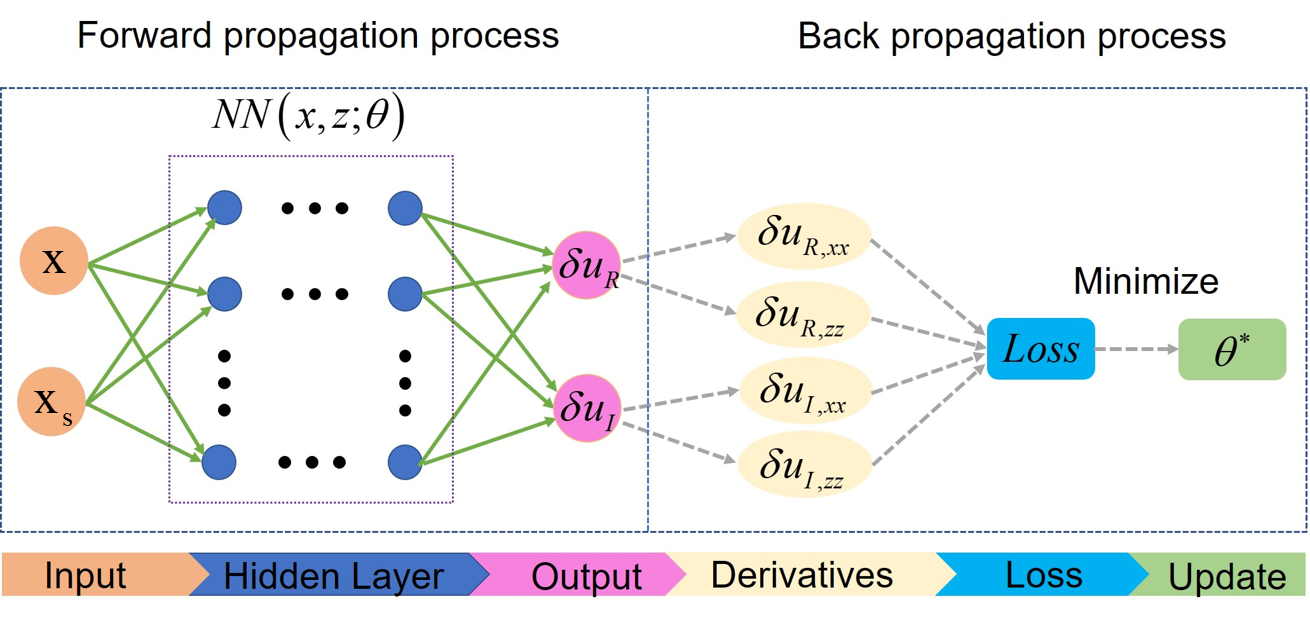}
\caption{The schematic of a PINN with inputs (spatial location $\bm{\mathrm{x}} =(x,z)$ and source location $\bm{\mathrm{x_s}} =(x_s,z_s)$) and outputs given by the real ($\delta u_R$) and imaginary $\delta u_I$ values of the complex scattered wavefield. The left-side dashed box illustrates a forward propagation process, and the right-side dashed box presents a back propagation process, in which derivatives ($\delta u_{R,xx}$, $\delta u_{R,zz}$, $\delta u_{I,xx}$, and $\delta u_{I,zz}$) are automatically calculated to update the PINN parameters ${\boldsymbol{\theta}}$.}
\label{fig1}
\end{figure}

Figure \ref{fig1} shows a schematic of PINN for solving for the scattered wavefield. As seen, we employ a fully connected NN, which consists of input, hidden, and output layers, with sine activation functions in every neuron, other than the ones in the last hidden layer. The output of PINN includes the real ($\delta u_{R}(\bm{\mathrm{x}},\omega,\bm{\mathrm{x_s}} )$) and imaginary ($\delta u_I(\bm{\mathrm{x}},\omega,\bm{\mathrm{x_s}} )$) parts of the complex scattered wavefield at the input coordinate location $(\bm{\mathrm{x}})$ and for a source at position $(\bm{\mathrm{x_s}})$. The training of PINN is constrained by a physical loss, which is determined by the governing equation \ref{eq2}, as follows:
\begin{equation}\label{eq4}
\begin{split}
\setlength{\abovedisplayskip}{3pt}
\setlength{\belowdisplayskip}{3pt}
\mathcal{L}_{p} = \frac{1}{N}\sum\limits_{j=1}^{N} \left ( {\left|(\omega^2 m^j + \nabla^2)\delta u_R^j + \omega^2 \delta m^j u_{R0}^j \right|_2^2 } \right. 
\left. {+ \left|(\omega^2 m^j + \nabla^2)\delta u_I^j + \omega^2 \delta m^j u_{I0}^j \right|_2^2 }\right),
\end{split}
\end{equation}
where $N$ represents the number of training samples derived from random selections within the 2D space domain, and $j$ is the training sample index. Here, for the sake of brivity, we focus the functional $(\bm{\mathrm{x}},\omega,\bm{\mathrm{x_s}})$ on $(\bm{\mathrm{x}})$ hereafter. In the loss function, the second-order partial derivatives of the scattered wavefield ($\delta u_{R}$ and $\delta u_{I}$) with respect to the spatial coordinates of the input are computed by automatic differentiation. The real ($u_{R0}$) and imaginary ($u_{I0}$) parts of the background wavefield can be quickly evaluated at any random spatial location analytically.

As we can see from equation \ref{eq4}, the velocity is a key parameter that controls the solution. Though the source location controls the center location of the wavefield, the velocity controls its shape. A change in the velocity model cause induce a large change in the wavefield shape, and thus, we often need to retrain the network to provide a solution for the new velocity model. However, such independent repeated training from random initialization leads to significant cost. Therefore, it is worthwhile to consider how to utilize past experience to guide the optimization of a subsequent PINN. 

\subsection{Meta-PINN}
In recent years, meta learning (MetaL) has received considerable attention in the machine learning community \cite{finn2017model}. It enables the network to learn how to learn and adapt to new tasks and domains efficiently. Hence, it can leverage prior knowledge and experience to learn new tasks efficiently. Here, we regard the seismic modeling for each velocity model as an independent task. Consequently, applying PINNs to obtain wavefield solutions for different velocity models is transformed into a multi-task learning problem. We utilize a MetaL algorithm to learn a task-level nonlinear mapping relationship, which can expedite the convergence of PINN training for new velocity models and enhance the accuracy of the wavefield solutions. The steps involved in the Meta-PINN algorithm is presented in Algorithm 1. Next, we will provide a detailed explanation of this algorithm and the steps involved.

We consider a PINN represented by a parameterized function ${G_{\boldsymbol{\theta}}}$ with learned parameters ${\boldsymbol{\theta}}$. When we adapt PINN to a new wavefield simulation task ${\mathcal{T}_i}$, i.e., solving the wave equation for a new velocity model sampled from the support data set (can be understood as the usual training data set), the PINN's parameters ${\boldsymbol{\theta}}$ are updated to ${\boldsymbol{\theta}}_i^{'}$. This update is accomplished through gradient descent methods over single or multiple iterations. For example, one gradient update is given by
\begin{equation}\label{eq5}
\setlength{\abovedisplayskip}{3pt}
\setlength{\belowdisplayskip}{3pt}
\boldsymbol{\theta}_i^{'} = {\boldsymbol{\theta}} - lr_{inner} \cdot \nabla_{\boldsymbol{\theta}} \mathcal{L}_{\mathcal{T}_i} \left( G_{\boldsymbol{\theta}} \right),
\end{equation}
where ${lr_{inner}}$ is the learning rate for inner iterations as a hyperparameter, and ${\mathcal{L}}$ denotes the loss function, which will be illustrated in the following section. 

Then, we evaluate the performance of the updated parameter $\theta_i^{'}$ on new velocity model. The new model is sampled from the query data set, which can be regarded as testing data set. Both support and query data sets contain the same number of velocity models, but correspond to a different distribution. After traversing all the velocity models in the training and test data set, we add the loss function values evaluated on all the samples in the query data set to optimize the PINN model parameters as follows
\begin{equation}\label{eq6}
\setlength{\abovedisplayskip}{3pt}
\setlength{\belowdisplayskip}{3pt}
{\boldsymbol{\theta}} \leftarrow {\boldsymbol{\theta}} - lr_{meta} \cdot \nabla_{\boldsymbol{\theta}} \sum_{\mathcal{T}_{i} \sim v (\mathcal{T})}^{} \mathcal{L}_{\mathcal{T}_i} ( G_{{\boldsymbol{\theta}}_i^{'}})
\end{equation}
where ${lr_{meta}}$ is meta (outer iterations) learning rate. The entire process described above constitutes a single epoch of training in the meta-training phase. By repeating these operations to a preset number of iterations, we end up with an NN model ${G_{\boldsymbol{\theta}}}$ that can be used to initialize PINN to predict wavefields for any new velocity model, efficiently. 

Here, we need to clarify two concepts to avoid confusion: updating network parameters and updating the network itself. In equation \ref{eq5}, what we update are the parameters of the network. This is done by copying the parameters of the Meta-PINN model from the current epoch, then using these copied parameters for forward propagation on the support dataset and calculating gradients, which are utilized to update the copied parameters. At this stage, we do not update the network model via backpropagation. Instead, we use the updated parameters for forward propagation directly on the query dataset, followed by calculating the loss and gradients on the query dataset. Only after traversing all velocity model distributions in the support and query datasets and accumulating the gradients from all query datasets, we use backpropagation to update the network model, as illustrated in equation \ref{eq6}. This procedure enables a bilevel gradient descent update from the support to the query sets. The aim is to ensure that the network parameters updated on the support set adapt effectively to the velocity models on the query set.

Once we train a Meta-PINN model, we can use it to provide the initialization for conventional training of a PINN network to obtain wavefield solutions for any velocity model. We refer to this process as meta-testing, which follows the same procedure as conventional PINN training. The only difference is that while conventional PINNs are typically randomly initialized, here we provide a robust initialization for the PINN, enabling it to converge rapidly to an accurate wavefield solution, as will be demonstrated in the subsequent numerical examples. 

\begin{algorithm}
\caption{Meta-PINN}\label{alg:Framwork}
\textbf{Input:} ${v(\mathcal{T})}$: Velocity distributions over tasks. \\
\textbf{Input:} ${lr_{inner}, lr_{meta}}$: Learning rate for inner and outer loops, respectively. \\
\textbf{Input:} ${iter}$: The number of iterations in the support dataset for every task. \\
\textbf{-------------------------------- Meta-training stage -----------------------------} \\
\textbf{Output:} Meta-based initialization of the PINNs model 
\begin{algorithmic}
\State 1: Randomly initialize PINN parameters ${\boldsymbol{\theta}}$
\State 2: \textbf{while} all tasks ${v(\mathcal{T})}$ \textbf{do}
\State 3: \quad Sample batch of tasks ${\mathcal{T}_i \sim v ( \mathcal{T})}$
\State 4: \quad \textbf{for} every $\mathcal{T}_i$ \textbf{do}
\State 5: \quad \quad \textbf{for} ${i}$ \textbf{in} ${iter}$ \textbf{do}
\State 6: \quad \quad \quad Evaluate $\nabla_{\boldsymbol{\theta}} \mathcal{L}_{\mathcal{T}_i} \left( G_{\boldsymbol{\theta}} \right)$ with respect to the support dataset for the sample \\ \quad \quad \quad \quad \quad \quad task $\mathcal{T}_i$
\State 7: \quad \quad \quad Compute adapted parameters with gradient descent: \\
\quad \quad \quad \quad \quad \quad \quad \quad \quad ${\boldsymbol{\theta}}_i^{'} = {\boldsymbol{\theta}} - lr_{inner} \cdot \nabla_{\boldsymbol{\theta}} \mathcal{L}_{\mathcal{T}_i} \left( G_{\boldsymbol{{\boldsymbol{\theta}}}} \right)$
\State 8: \quad \quad \textbf{end for}
\State 9: \quad \quad Evaluate $ \mathcal{L}_{\mathcal{T}_i}( G_{{\boldsymbol{\theta}}_i^{'}})$ with respect to the query dataset from the sample task $\mathcal{T}_i$
\State 10: \quad \textbf{end for}
\State 11: \quad Sum the loss of all tasks on the query dataset: $\mathcal{L}_{sum} = \sum_{\mathcal{T}_i \sim v \left( \mathcal{T} \right)} \mathcal{L}_{\mathcal{T}_i} ( G_{{\boldsymbol{\theta}}_i^{'}})$
\State 12: \quad Update the Meta-PINN ${\boldsymbol{\theta}} \leftarrow {\boldsymbol{\theta}} - lr_{meta} \cdot \nabla_{\boldsymbol{\theta}} \mathcal{L}_{sum}$
\State 13: \textbf{end while}
\State 14: \textbf{Return:} Meta-PINN parameters ${\boldsymbol{\theta}}$
\end{algorithmic}
\textbf{-------------------------------- Meta-testing stage -----------------------------}\\
\textbf{Output:} Velocity-specific PINN model 
\begin{algorithmic}
\State 15: Fine-tune the Meta-PINN parameters ${\boldsymbol{\theta}}$ on each specific velocity model
\State 16: Testing the updated model to obtain the wavefield solutions
\end{algorithmic}
\end{algorithm}

\subsection{Loss functions}
In PINNs, we incorporate physical constraints in the loss function, making it necessary for the loss function to not only fit the data but also ensure that the predicted solutions adhere to the underlying physical laws. In this case, the loss function plays a crucial role in the training of PINNs because it determines how well the NN is able to capture complex physical problems. In the context of wave physics, which is often a high-dimensional problem, we require careful consideration of the loss function used to train the PINNs. Although physical loss can provide a constraint on the training of PINNs, they can fail and sometimes lead to poor performance and convergence rates, as the model may converge to trivial solutions in the absence of effective hyperparameters and training strategies. As such, the optimization of PINNs for solving the wave equation must strike a delicate balance between incorporating physical constraints while also ensuring that the model is capable of accurately predicting the wavefield solutions. 

Therefore, in addition to employing a physical loss, we also utilize a regularization loss as follows \cite{huang2023gaborpinn}:
\begin{equation}\label{eq7}
\begin{split}
\setlength{\abovedisplayskip}{3pt}
\setlength{\belowdisplayskip}{3pt}
\mathcal{L}_{r} = \frac{1}{N_{reg}}\sum\limits_{i=1}^{N_{reg}}  \left[(\delta u_R^i)^2 + (\delta u_I^i)^2 \right],
\end{split}
\end{equation}
where ${N_{reg}}$ denotes the number of samples used to calculate the
penalty term, which is often taken from around the source location. The physical reason behind reducing the scattered wavefield energy around the source and in general, is that most of the wavefield, especially around the source, is represented by the background wavefield. This is especially the case if the homogenoues velocity used to calculate the background wavefield is the velocity at the source location. This loss has been validated to stabilize the training of PINNs and to prevent trivial solutions.

In both meta training and meta testing stages, the total loss of the network is expressed as
\begin{equation}\label{eq8}
\begin{split}
\setlength{\abovedisplayskip}{3pt}
\setlength{\belowdisplayskip}{3pt}
\mathcal{L} = \epsilon \cdot (\alpha_1 \cdot \mathcal{L}_{p} + \alpha_2 \cdot \mathcal{L}_{r}),
\end{split}
\end{equation}
where $\mathcal{L}_{p}$ is defined in equation \ref{eq4}, the hyperparameters $\alpha_1$ and $\alpha_2$ are used to balance the physical and regularization losses, and $\epsilon$ represents a scaling factor that is employed to adjust the magnitude of the total loss value. In our implementation, for simplification, the hyperparameters $\alpha_1$ and $\alpha_2$ are set to 1. For the value of parameter $\epsilon$, to maintain meta-training stability, we set $\epsilon = 0.1$ during the meta-training phase. In the meta-testing phase, $\epsilon$ remains at 0.1 to ensure consistency with the configuration during the meta-training phase.
\section{Numerical Examples}
\subsection{Meta-training procedure}
In the following tests, we consider training PINNs to learn frequency domain wavefields of 5 Hz, solutions of equation \ref{eq2}. As previously stated, our Meta-PINN algorithm requires an initial meta-training phase to provide a robust initialization. For this, we create 20 unique velocity models for the meta-training phase, each with distinct structures and velocity distribution ranges. Half of these models are used as the support set, and the other half as the query set. All velocity models have a velocity distribution ranging of $1.5\sim5$ km/s, with the smallest model having a spatial size of $2 \text{km} \times 2 \text{km}$, and the largest $5 \text{km} \times 20 \text{km}$. we draw 40000 spatially random samples for training, and the samples include their spatial coordinates $x$ and $z$, the corresponding source's location $\mathrm{x_s}$, as well as the velocity $v$, and the background velocity $v_0$ at the random drawn location. Here, the source is placed near the surface, with a fixed depth of 0.025 km.

We employ a multi-layer perceptron (MLP) with six hidden layers as our baseline network, and the neurons in each of the six hidden layers are $\{256, 256, 128, 128, 64, 64\}$ from shallow to deep. The network undergoes a total of 50000 training epochs in the meta-training phase using an AdamW optimizer \cite{loshchilov2017decoupled}. The number of inner loop iterations is set to 20, with a outer loop learning rate of 1e-3, decreasing to 0.8 of its value every 5000 epochs. The inner loop learning rate is consistently maintained at 2e-3.

After obtaining the meta-learning-based network initialization, we first test it in a simple layered model, and then share the test results in the more realistic Overthrust model. To demonstrate the effectiveness of the proposed Meta-PINN algorithm, we compare the convergence speed and accuracy of the meta-initialized PINN (denoted as Meta-PINN) with that of the vanilla PINN (i.e., randomly initialized).

\subsection{A layered model}
\begin{figure*}[!t]
\centering
\includegraphics[width=0.32\textwidth]{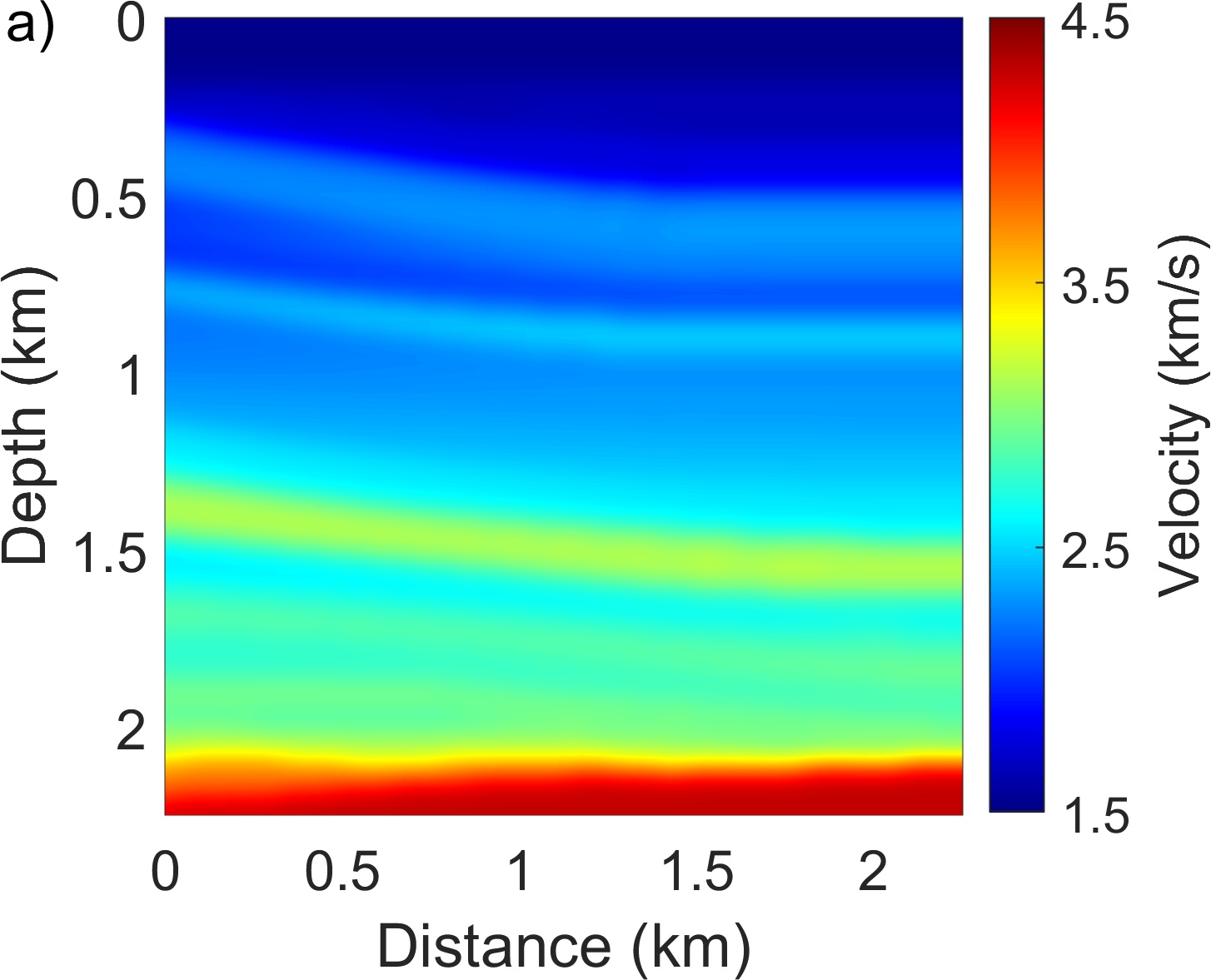}
\includegraphics[width=0.32\textwidth]{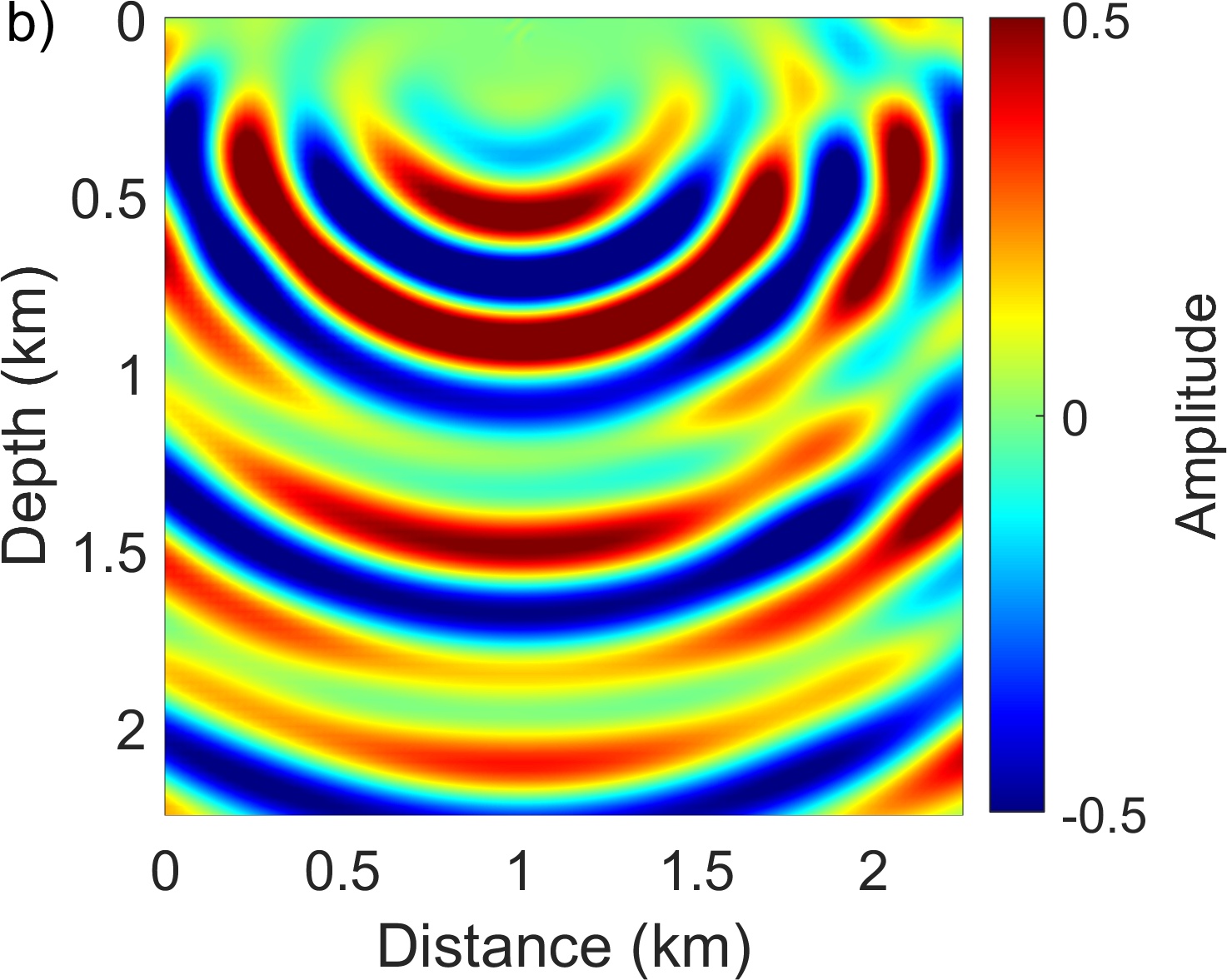}
\includegraphics[width=0.32\textwidth]{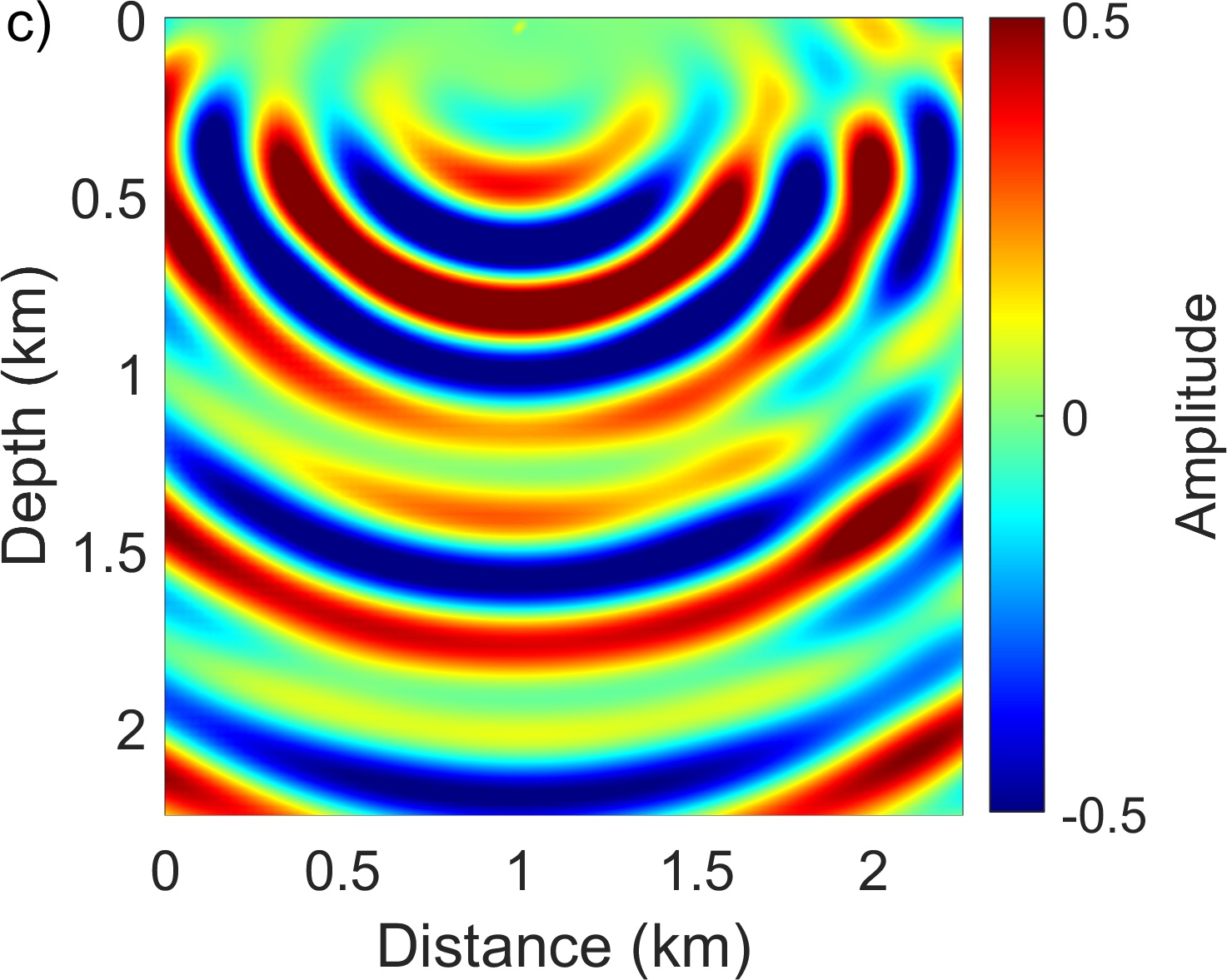}
\caption{The velocity model (a) and the real (b) and imaginary parts (c) of the 5 Hz scattered wavefield calculated numerically. The background velocity is constant and equal to 1.5 km/s. }
\label{fig2}
\end{figure*}

Figure \ref{fig2}a displays the tested simple layered model, measuring $2.25 \text{km} \times 2.25 \text{km}$, extracted from the Marmousi model. Using the FDM, we numerically solved Equation \ref{eq2} to obtain the real and imaginary parts of the 5 Hz scattered wavefield, which are used as references for the subsequent neural network wavefield solutions, depicted respectively in Figures \ref{fig2}b and \ref{fig2}c. The background velocity is constant at 1.5 km/s, with the source located at the surface at $x=1$ km. For this test model, as with the meta-learning training, we utilize 40000 random samples for training both the vanilla PINN and Meta-PINN. The training of both networks use the AdamW optimizer, with an initial learning rate of 1.2e-3, halved at the 2000th epoch.

\begin{figure*}[!t]
\centering
\includegraphics[width=0.5\textwidth]{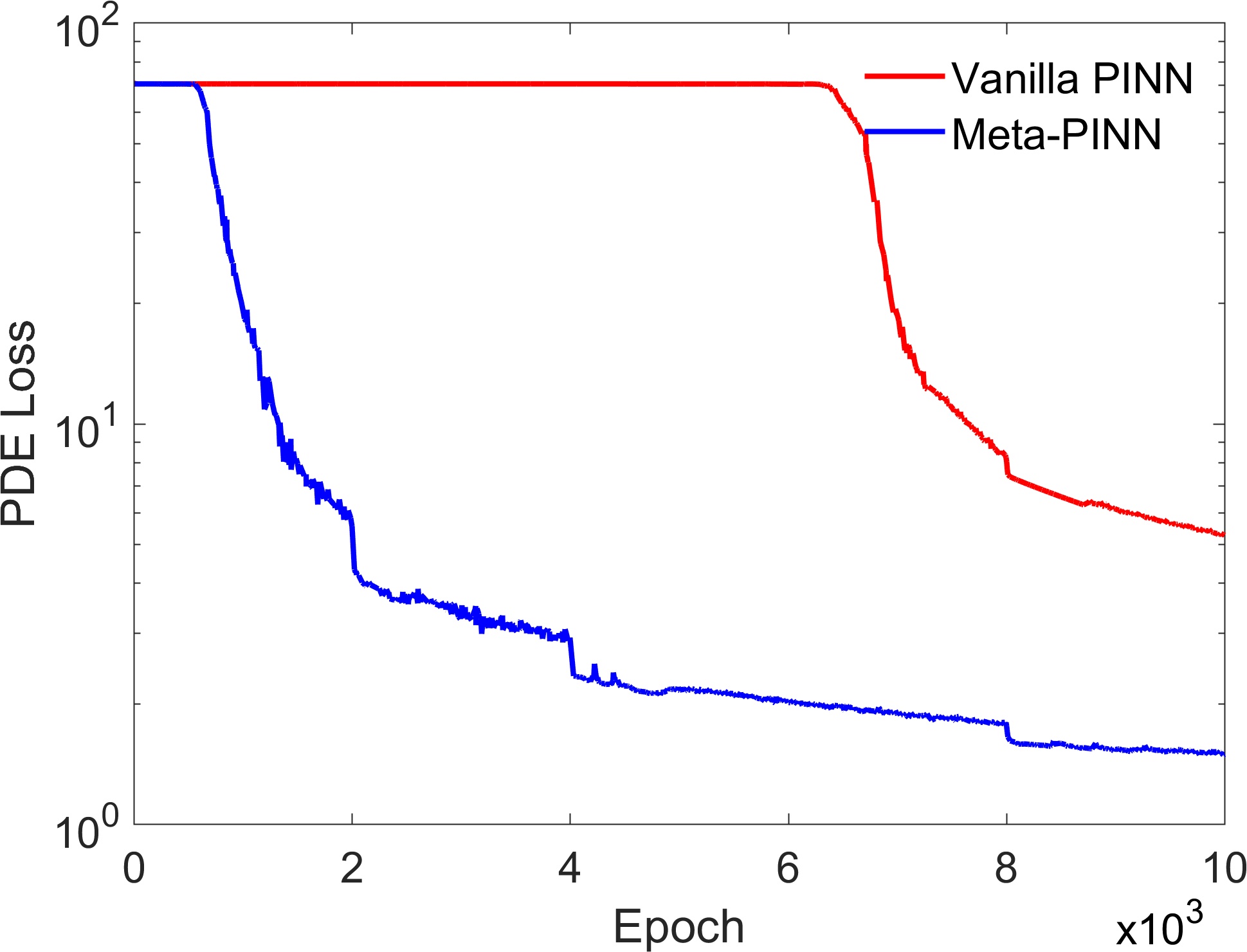}
\caption{A comparison of the physical loss curves between the vanilla PINN and Meta-PINN for the simple layered model (shown in Figure \ref{fig2}a). }
\label{fig3}
\end{figure*}

\begin{figure*}[!t]
\centering
\includegraphics[width=1\textwidth]{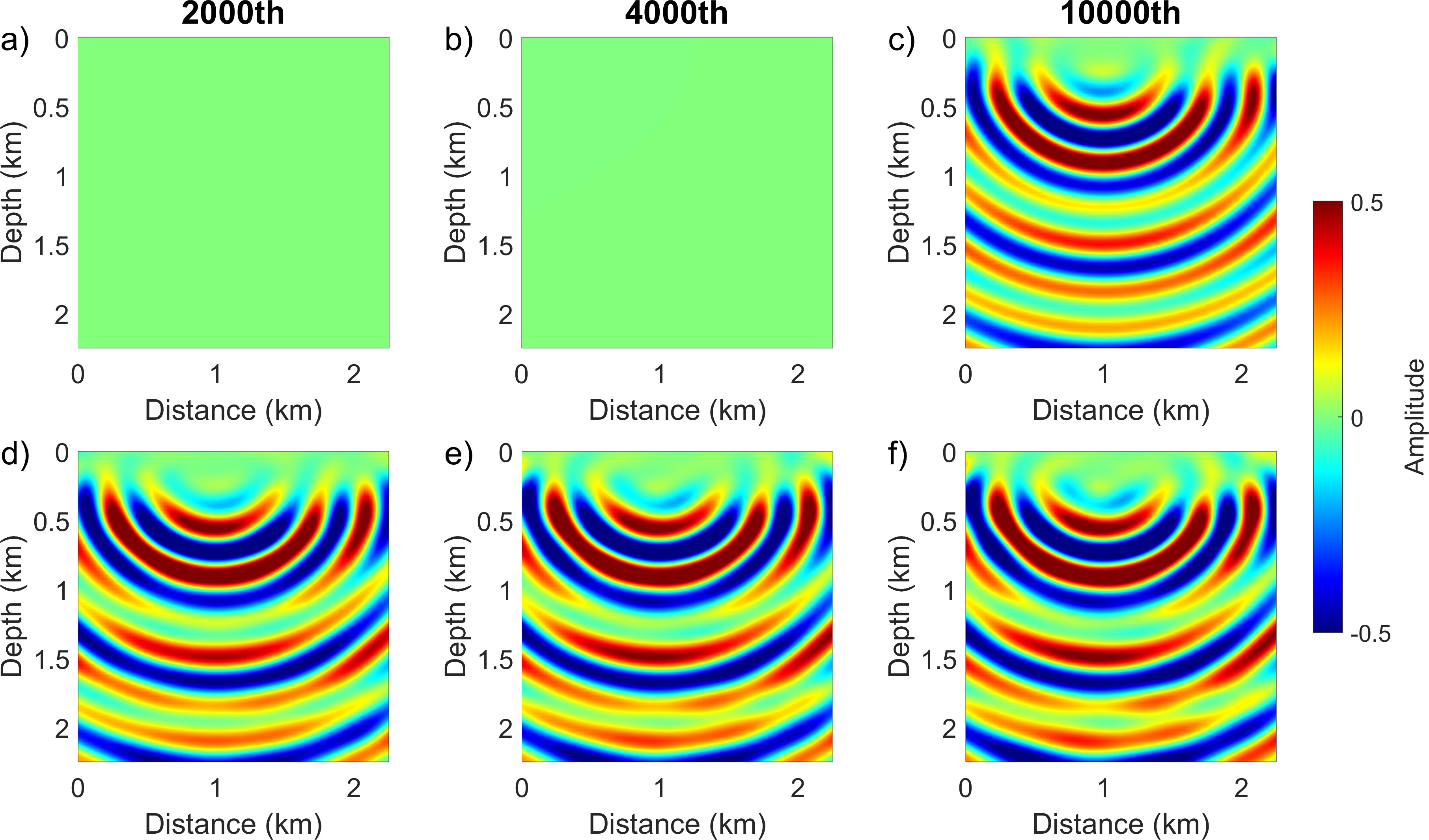}
\caption{The predicted real part of the scattered wavefield from the vanilla PINN and Meta-PINN after 2000 (fist column), 4000 (second column), and 10000 (third column) epochs of training for the layered model. The first and second rows correspond to the predicted scattered wavefield of the vanilla PINN and Meta-PINN, respectively.}
\label{fig4}
\end{figure*}

\begin{figure*}[!t]
\centering
\includegraphics[width=1\textwidth]{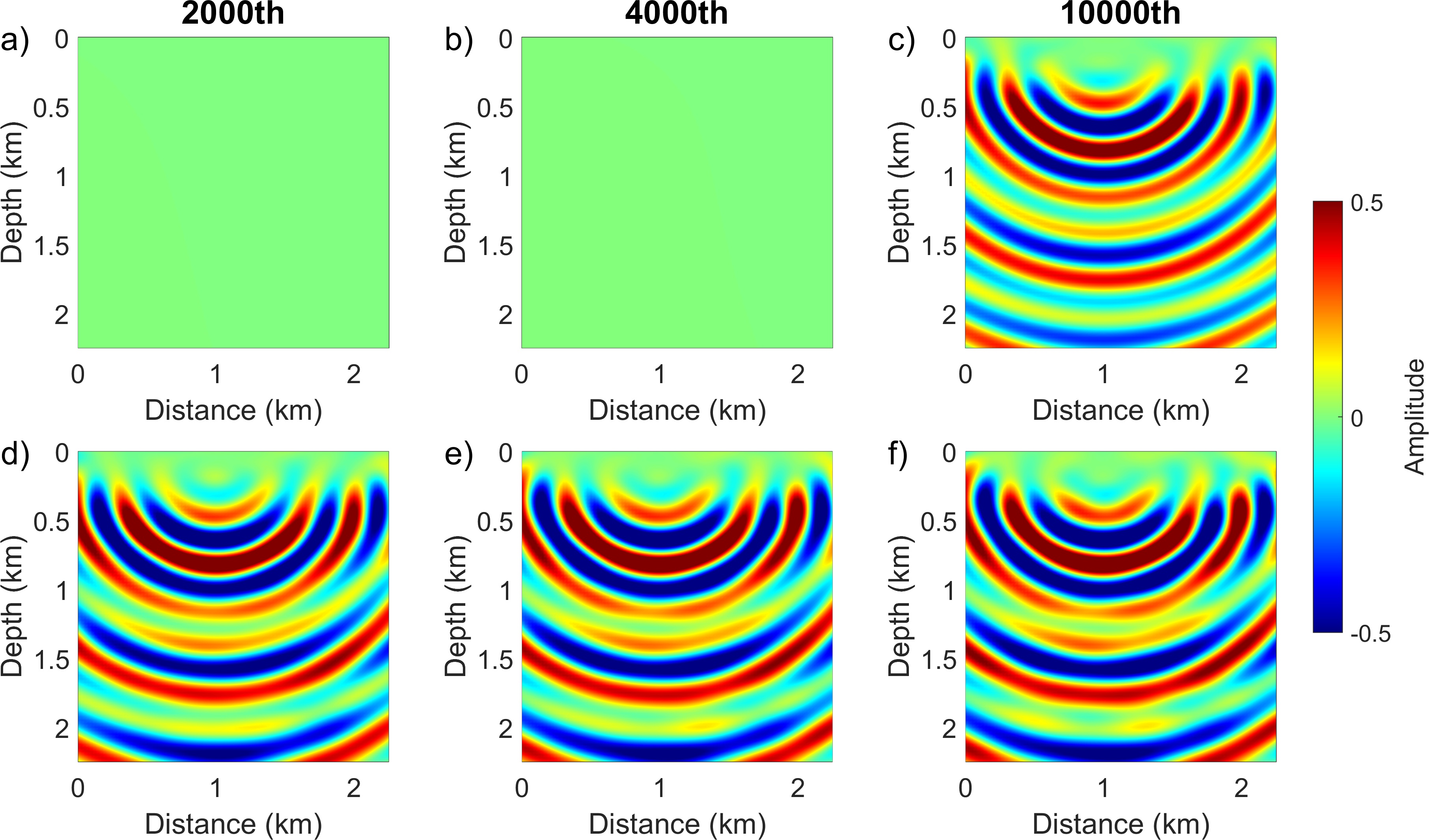}
\caption{Similar with Figure \ref{fig4}, but for the predicted imaginary part of the scattered wavefield.}
\label{fig5}
\end{figure*}

Figure \ref{fig3} presents a comparison of the convergence speeds of the physical loss between the vanilla PINN and Meta-PINN. As we can see in the figure, Meta-PINN converges much faster than the vanilla PINN. The vanilla PINN spends a long time at the beginning exploring the parameter space without convergence until it manages to latch to a local minimum. Numerically, our Meta-PINN achieves the accuracy of a vanilla PINN trained for 10000 epochs in approximately just 2000 epochs. We further compare the real and imaginary parts of the scattered wavefield predicted by both networks, as shown in Figures \ref{fig4} and \ref{fig5}, respectively. It is evident that Meta-PINN provides a high-accuracy wavefield solution with only 2000 epochs of training (Figures \ref{fig4}d and \ref{fig5}d), with the wavefield undergoing only minor refinements as training progresses (Figures \ref{fig4}e-f and \ref{fig5}e-f). In contrast, the vanilla PINN fails to learn the appropriate network parameters for predicting the wavefield solution at the 2000th and 4000th epochs (Figures \ref{fig4}a-b and \ref{fig5}a-b). Even after 10000 epochs of training, the scattered wavefield predicted by the vanilla PINN still lacks sufficient details (Figures \ref{fig4}c and \ref{fig5}c).

\subsection{Overthrust model}
\begin{figure*}[!t]
\centering
\includegraphics[width=0.8\textwidth]{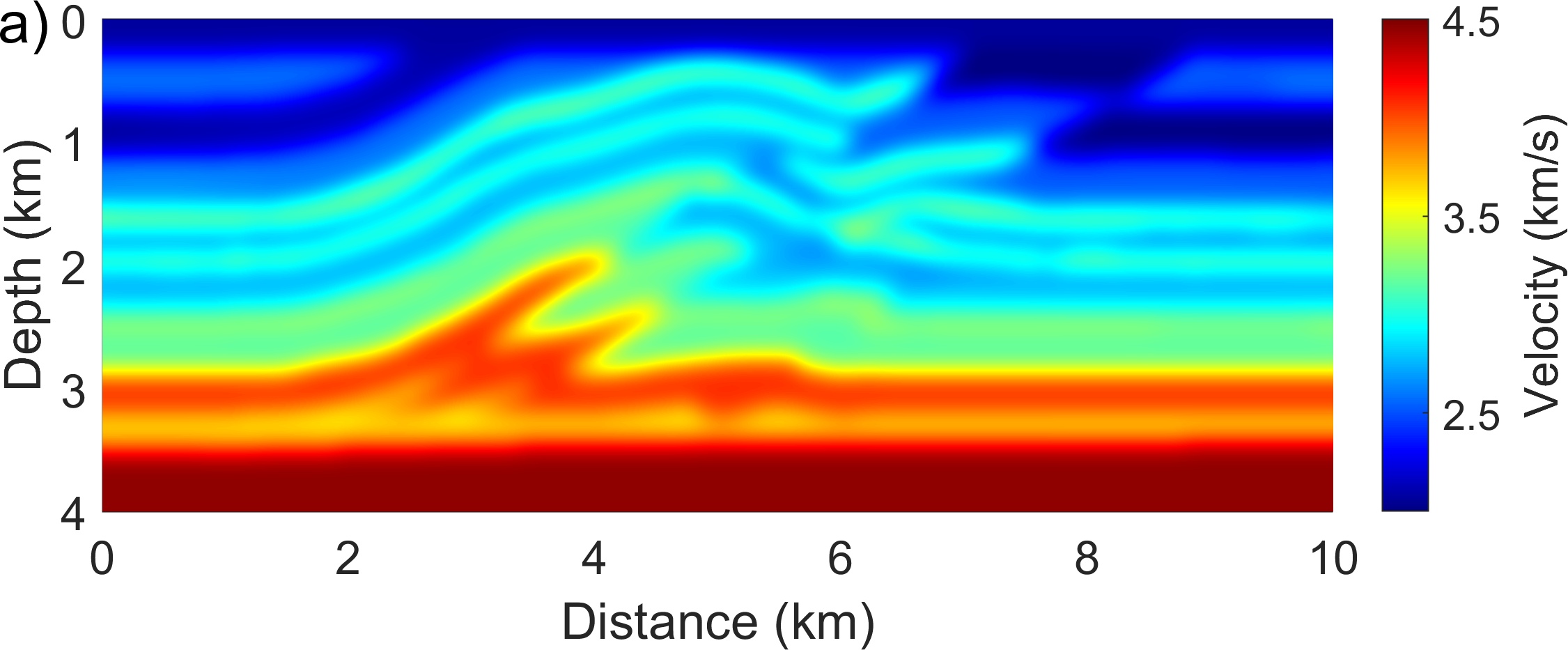}
\includegraphics[width=0.8\textwidth]{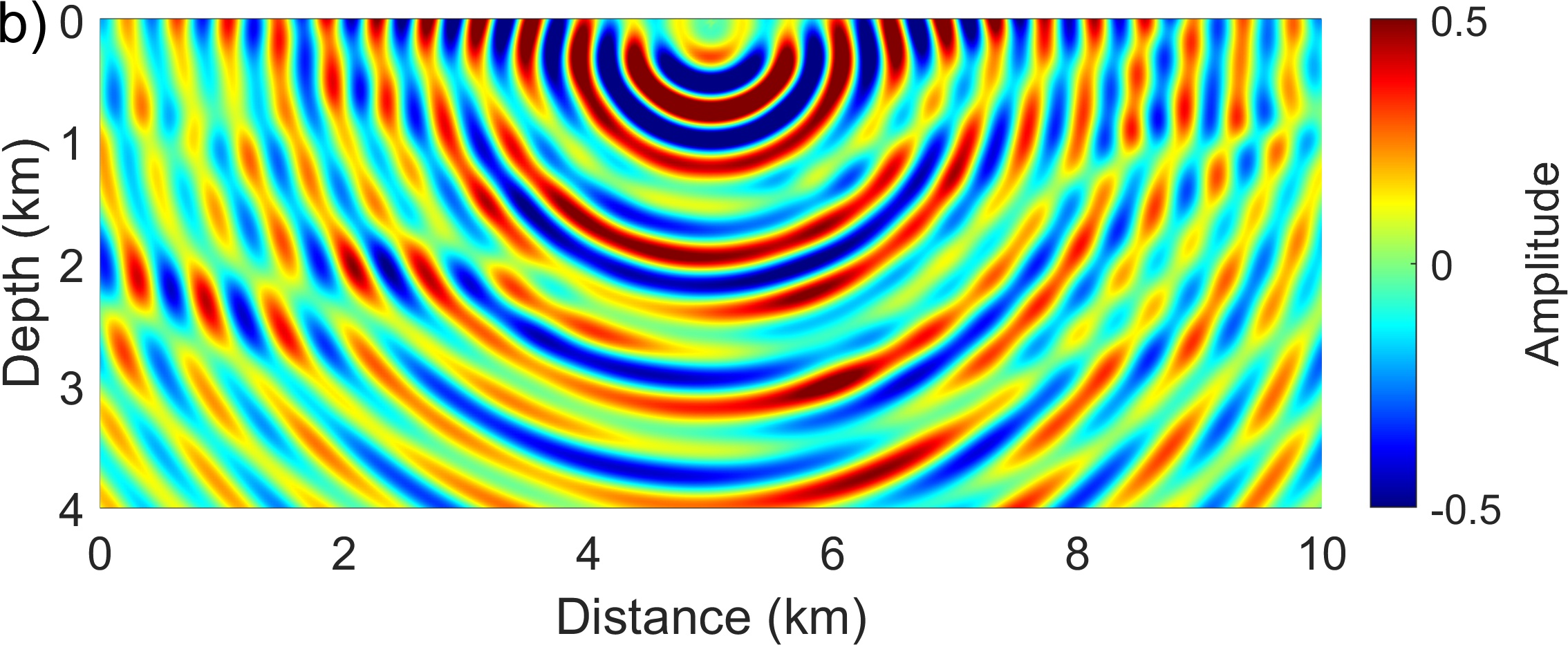}
\includegraphics[width=0.8\textwidth]{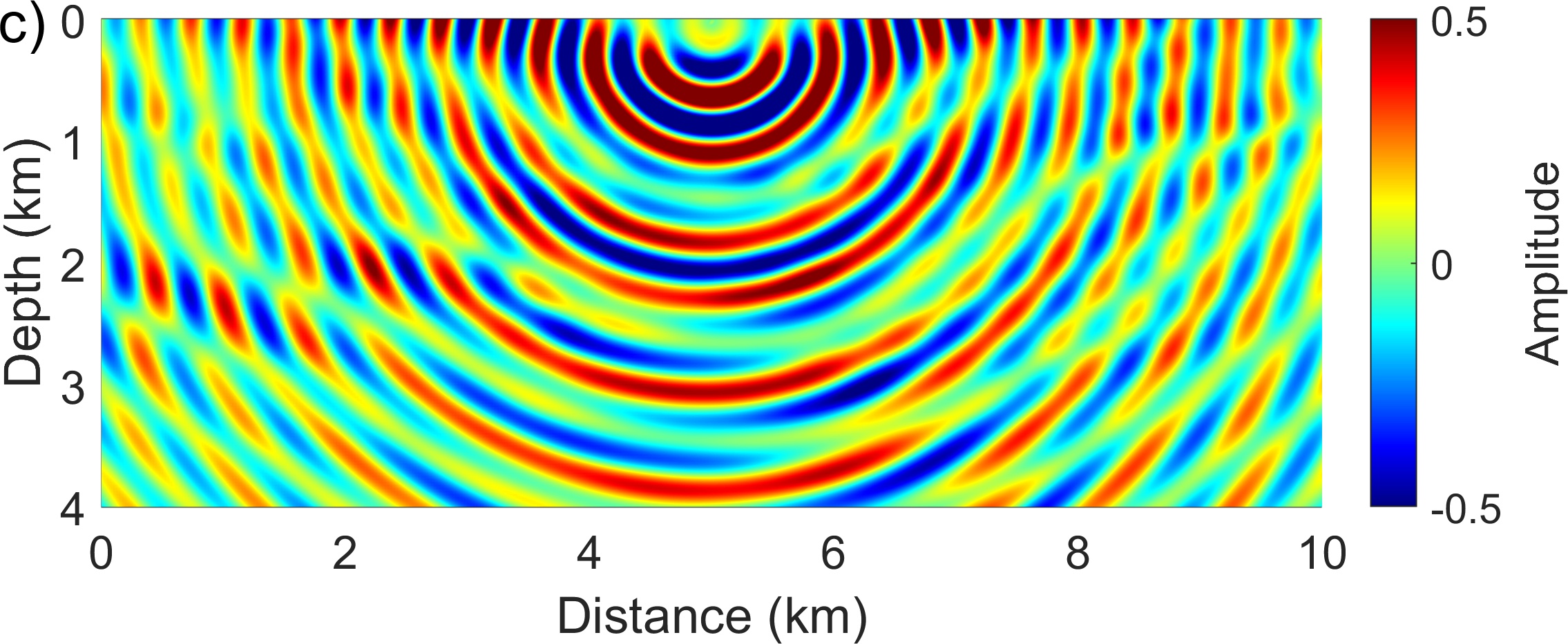}
\caption{The velocity model (a) and the real (b) and imaginary parts (c) of the 5 Hz scattered wavefield calculated numerically. The background velocity is constant and equal to 2 km/s. }
\label{fig6}
\end{figure*}

We, then, share our test results on a more complex overthrust model (see Figure \ref{fig6}a), a classic model in the seismic community. Compared to the previous layered model, the overthrust model has a larger size, measuring 10 km in width and 4 km in depth. Meanwhile, the overthrust model includes more layers and two complex overthrust faults in the middle. For evaluation, we again use the FDM to obtain a numerical reference solution. The source is located at the surface's center, and the background velocity is 2 km/s. Compared to the wavefield solution of the layered model in the previous section, the scattered wavefield of the overthrust model contains more details, as well as faults. We also use 40000 random samples for training both the vanilla PINN and Meta-PINN. The AdamW optimizer is used for training the networks, with an initial learning rate of 3e-3, which is decreased by a factor of 0.5 at the 2000th, 4000th, and 8000th epochs.

\begin{figure*}[!t]
\centering
\includegraphics[width=0.5\textwidth]{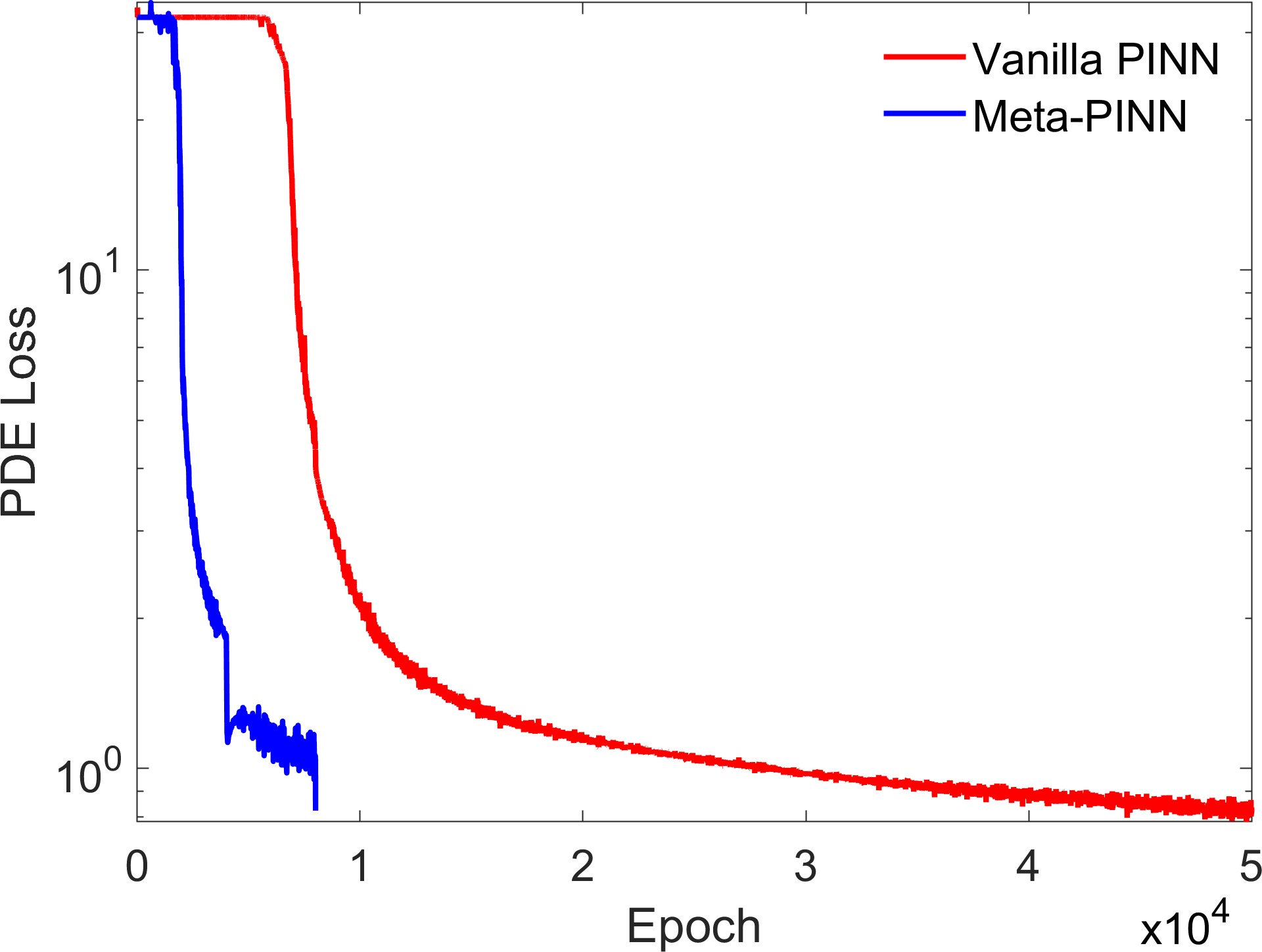}
\caption{A comparison of the physical loss curves between the vanilla PINN and Meta-PINN for the overthrust model (shown in Figure \ref{fig6}a).  }
\label{fig7}
\end{figure*}

\begin{figure*}[!t]
\centering
\includegraphics[width=1\textwidth]{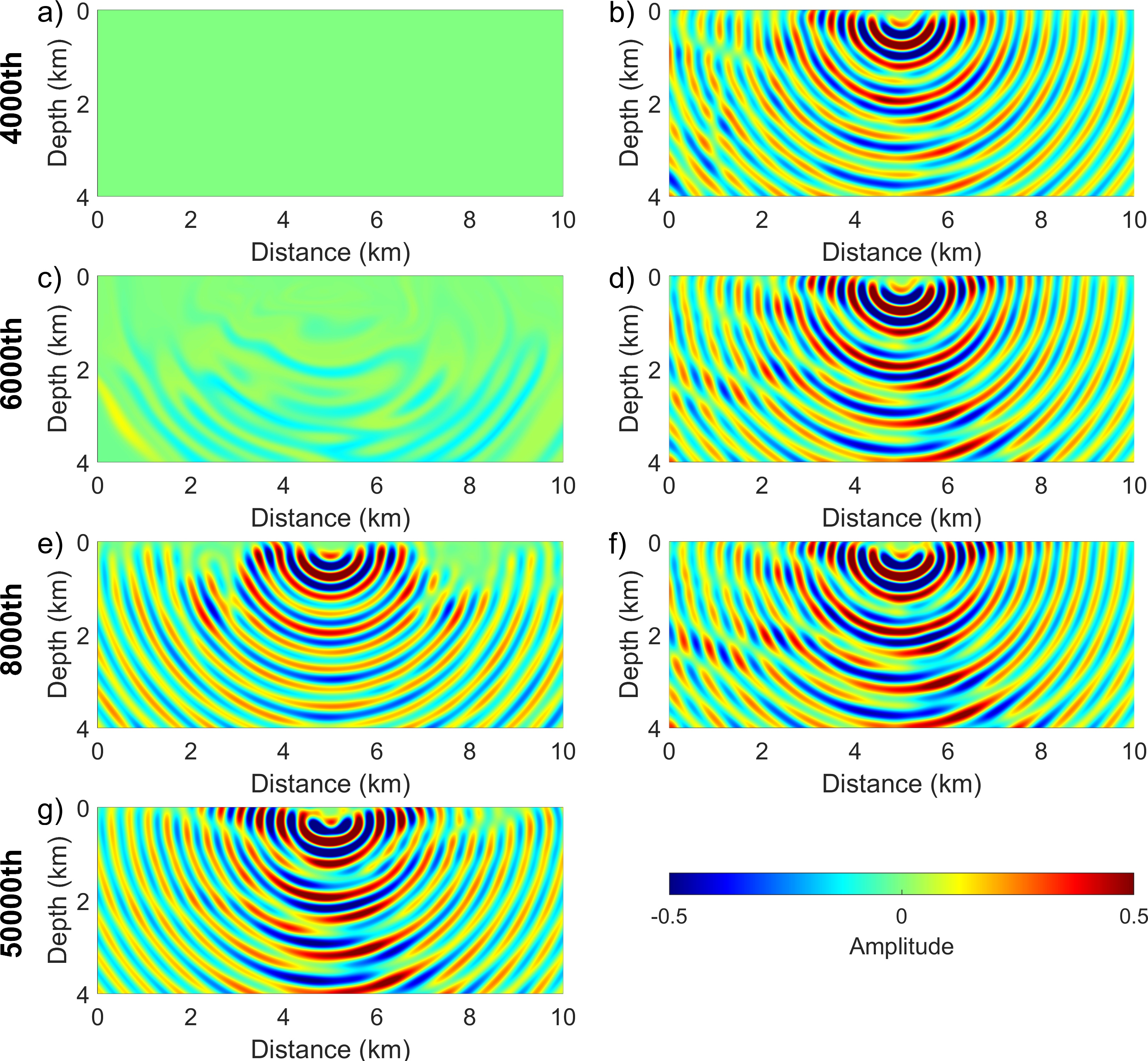}
\caption{The predicted real part of the scattered wavefield from the vanilla PINN and Meta-PINN after 4000 (fist row), 6000 (second row), 10000 (third row), and 50000 (fourth row) epochs of training for the overthrust model. The first and second columns correspond to the predicted scattered wavefield of vanilla PINN and Meta-PINN, respectively.}
\label{fig8}
\end{figure*}

\begin{figure*}[!t]
\centering
\includegraphics[width=1\textwidth]{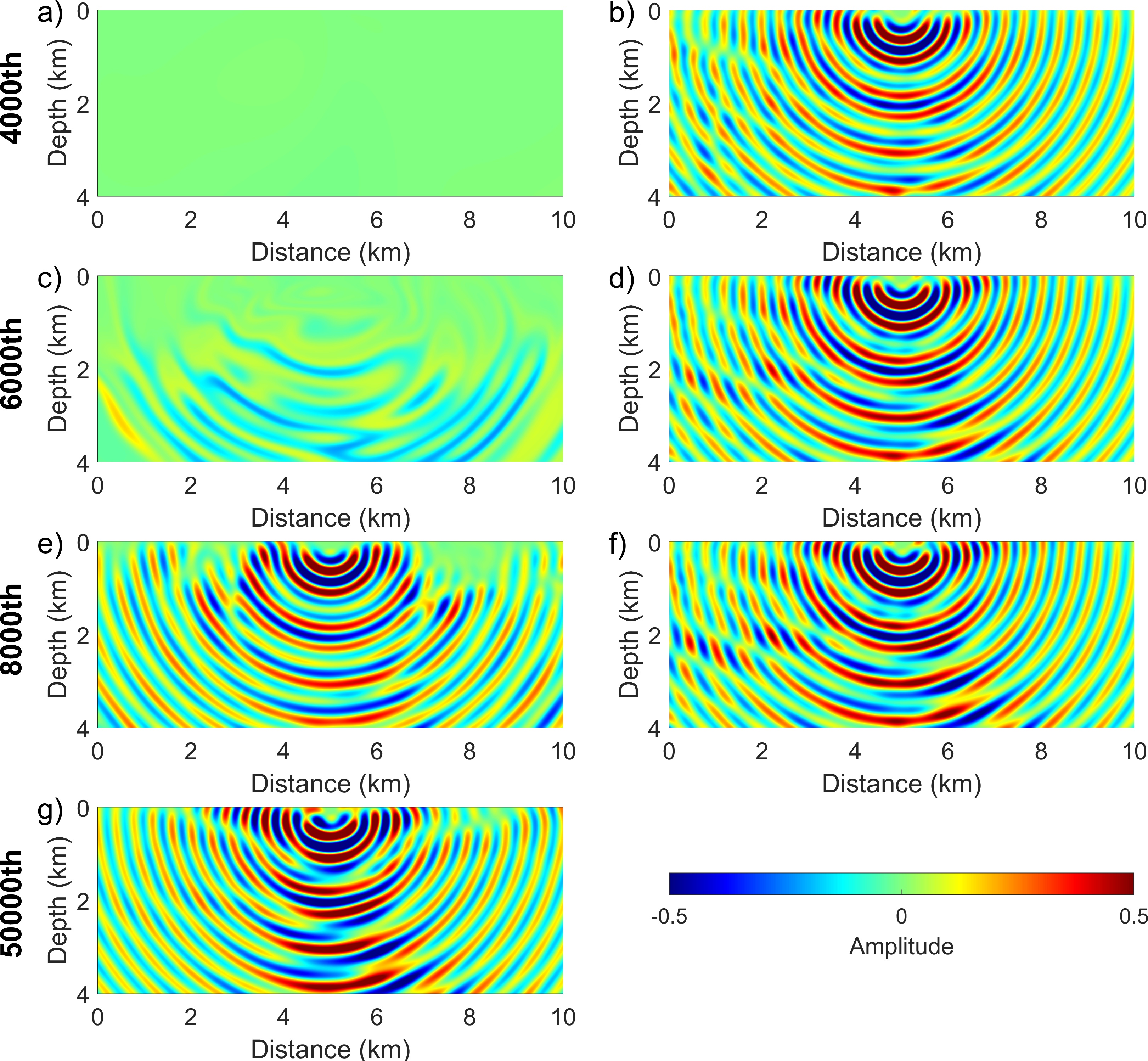}
\caption{Similar with Figure \ref{fig8}, but for the predicted imaginary part of the scattered wavefield.}
\label{fig9}
\end{figure*}

Figure \ref{fig7} displays the physical loss curves of both PINNs. Again, our Meta-PINN exhibits good convergence, particularly noticeable from the outset. In contrast, the vanilla PINN spends considerable time exploring the parameter space. Numerically, our Meta-PINN reaches the predefined loss threshold in about 8000 epochs, whereas the vanilla PINN requires training for 50000 epochs. We further compare the real and imaginary parts of the scattered wavefield solutions provided by both PINNs at different epochs, shown in Figures \ref{fig8} and \ref{fig9}. Once again, our Meta-PINN provides an acceptable wavefield solution by the 4000th epoch (Figures \ref{fig8}b and \ref{fig9}b). With prolonged training, the wavefield solution from the Meta-PINN becomes increasingly refined, offering more details (Figures \ref{fig8}d,f and \ref{fig9}d,f). Conversely, within the first 6000 epochs, the vanilla PINN fails to learn an effective wavefield representation (Figures \ref{fig8}a,c and \ref{fig9}a,c), achieving partial reconstruction only by the 8000th epoch (Figures \ref{fig8}e and \ref{fig9}e). Even after 50000 epochs, the wavefield solution from the vanilla PINN still shows significant discrepancies compared to the numerical reference solution (Figures \ref{fig8}g and \ref{fig9}g). These results demonstrate the superiority of our Meta-PINN algorithm in convergence speed and accuracy for solving complex model wavefields, compared to the vanilla PINN.

\section{Discussion}
In the following, we will share some of the advantages and limitations of the approach by addressing the following queries: 
\subsection{What will we obtain from meta-initialization?}
\begin{figure*}[!t]
\centering
\includegraphics[width=0.95\textwidth]{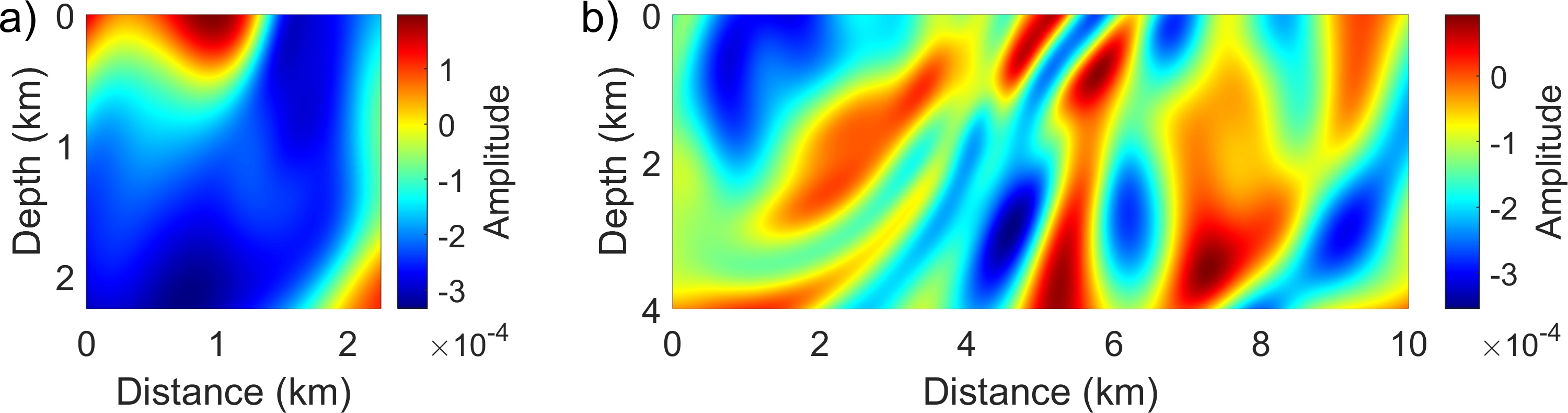}
\caption{The predicted real part of the scattered wavefield using the meta-initialized network parameters: (a) the layered model and (b) the overthrust model.}
\label{fig10}
\end{figure*}

As demonstrated in our numerical examples, the proposed Meta-PINN admitted outstanding improvements in convergence speed and prediction accuracy over the vanilla PINN. Essentially, as previously stated, the Meta-PINN and vanilla PINN differ only in their network initialization. This leads us to wonder, what kind of wavefield solution would we obtain if we directly predicted using the meta-initialized parameters (Epoch zero)? Figure \ref{fig10} shows the real part of the scattered wavefield solutions predicted directly using the meta-initialized network parameters for both the test layered and overthrust models. We can see that the meta-initialized parameters do not directly provide a reasonable wavefield solution. However, remarkably, these meta-initialized parameters enable the PINN to converge rapidly to the desired performance, providing an accurate wavefield solution. This is attributed to the core idea of meta-learning, which is not focused on the neural network's current state in a specific task, but on finding an initial state that allows the neural network to converge quickly after a minimal amount of gradient updates on new tasks. In our implementation, we train a model on a number of different velocity distributions, incorporating bidirectional gradient updates from the support to the query sets, enabling it to learn good initial network parameters. These parameters do not need to be optimal for solving the wavefield for a specific velocity distribution, but they need to be an excellent starting point for rapid adaptation to new velocity distributions.

\subsection{What amount of training data is needed for Meta-PINN?}
Unlike the vanilla PINN, our Meta-PINN algorithm requires a preliminary meta-training phase to meta-learn a robust initialization. During the meta-training stage, we construct two specific training sets, named the support and the query sets. The support set is used for updating network parameters (not the network itself) in the inner loop, and these updated parameters are then directly evaluated on the query set to obtain the loss for the outer loop. We traverse all velocity distributions in the support and query sets and accumulate all losses from the query set to update the network. Consequently, the number of training velocity models in the support and query sets may be a crucial factor influencing the robustness of the meta-initialized parameters.

\begin{figure*}[!t]
\centering
\includegraphics[width=0.5\textwidth]{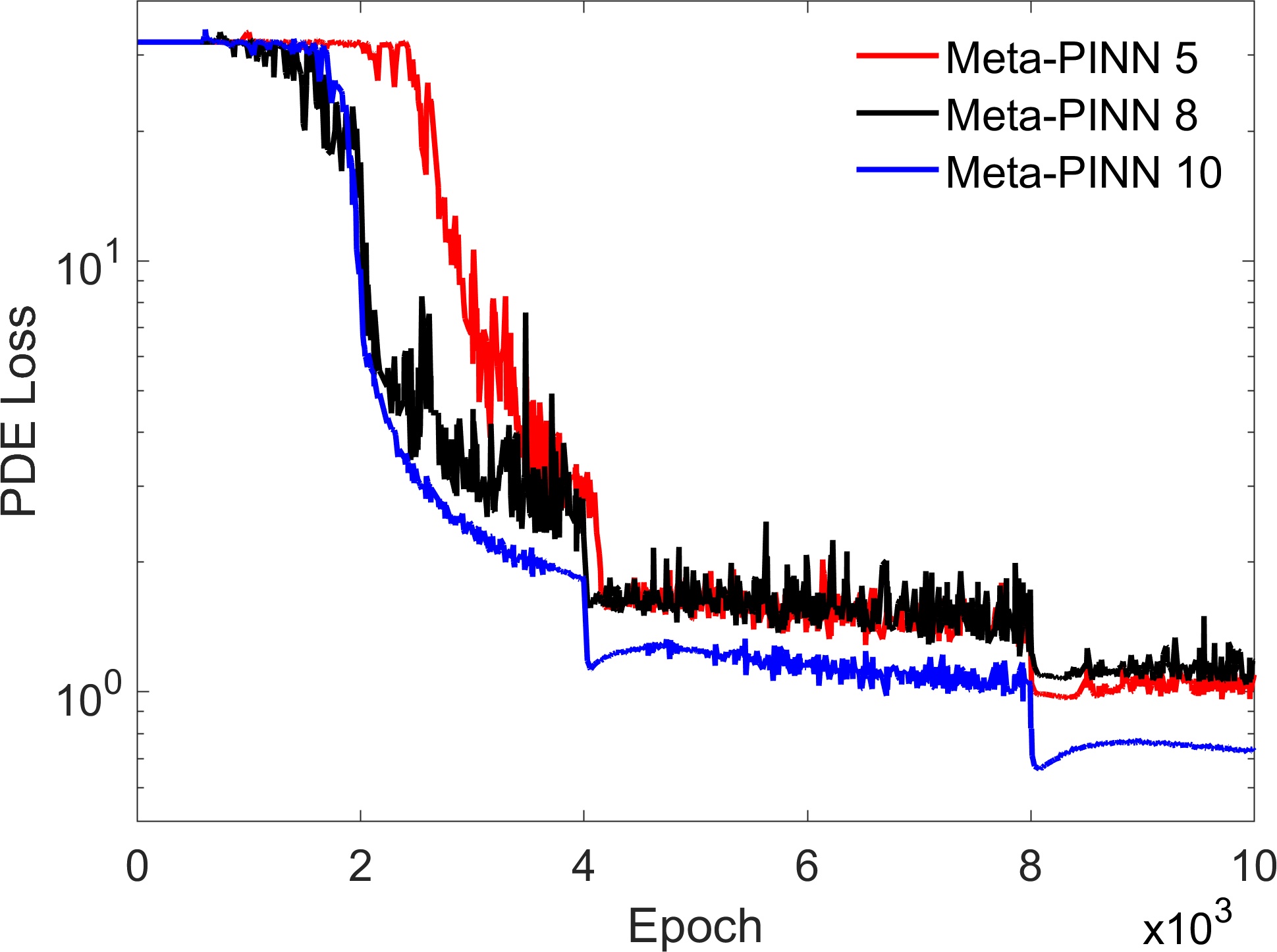}
\caption{The physical loss curves during the meta-testing phase for the meta-initialized parameters trained with different numbers of velocity models. 'Meta-PINN 5', 'Meta-PINN 8', and 'Meta-PINN 10' respectively represent the support set and query set of meta-training phase having 5, 8, and 10 velocity models each.}
\label{fig11}
\end{figure*}

So, we will conduct a test to analyze the impact of the amount of training data on the performance of the Meta-PINN method. In this test, we perform two more meta-trainings to train two different meta-initialized parameters. The first meta-training had 8 training samples each in its support and query sets, while the second had only 5 training samples each. The two meta-initialized network parameters trained will undergo meta-testing on the overthrust model, with the refined PINN networks respectively termed 'Meta-PINN 8' and 'Meta-PINN 5'. Their convergence curves are compared with the Meta-PINN trained in the numerical example section (indicated as Meta-PINN 10) and displayed in Figure \ref{fig11}. We observe that involving more velocity models in the meta-training phase enhances the performance of the meta-initialized parameters, enabling the PINN to converge faster and provide a more accurate wavefield solution. This is predictable since more training samples allows the network to capture a wider range of velocity distribution features during the meta-training phase, such as the velocity range, model size, and geological structure. However, we also acknowledge that a larger training set increases memory and training time during the meta-training phase. Therefore, to balance the training burden and performance, we chose to train the meta-initialized parameters with 10 velocity models.

\subsection{How many iterations we need for the inner loop?}
\begin{figure*}[!t]
\centering
\includegraphics[width=0.5\textwidth]{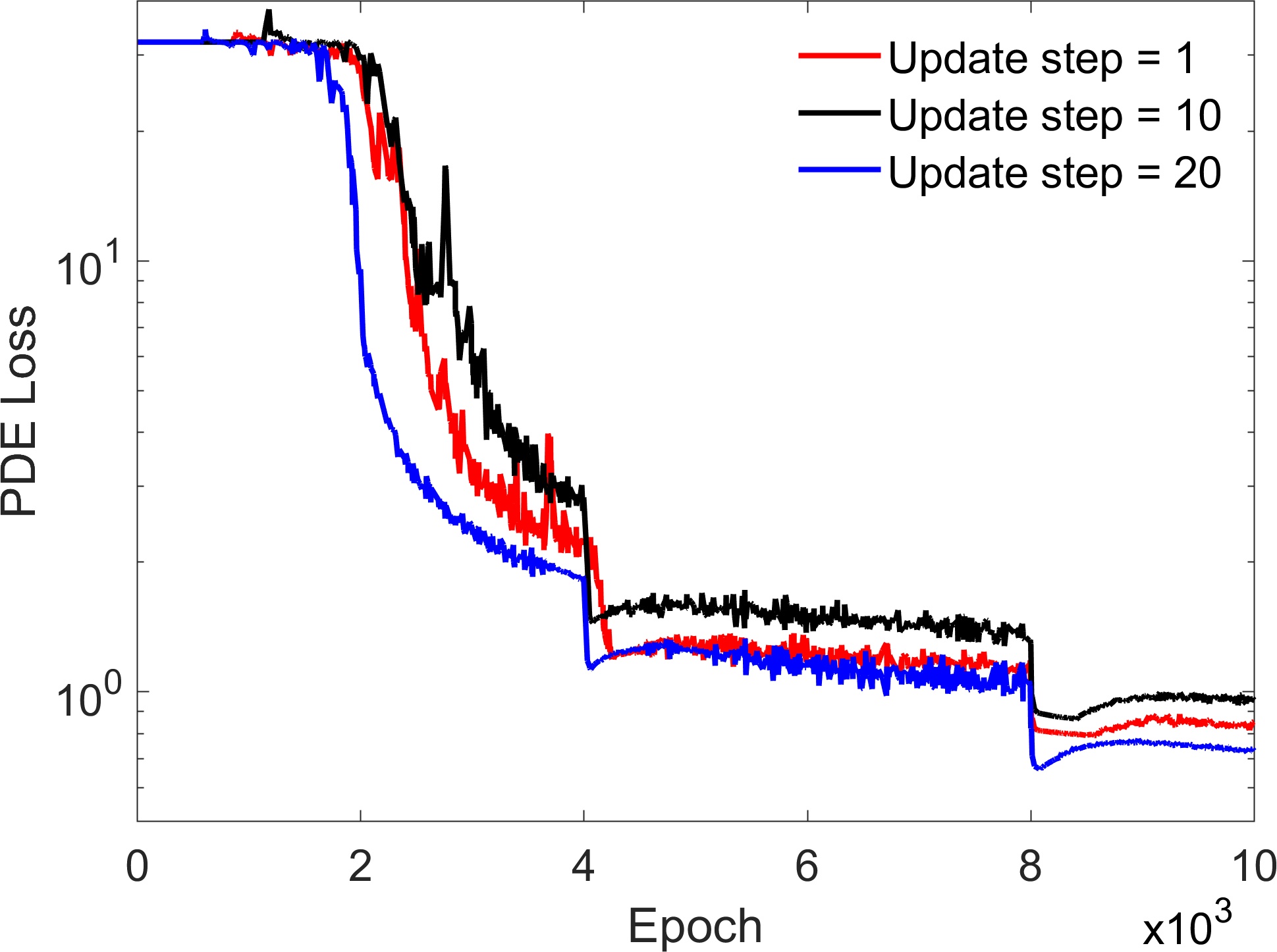}
\caption{The physical loss curves during the meta-testing phase for the meta-initialized parameters trained with different numbers of iterations in the inner loop. 'Update step = 1', 'Update step = 10', and 'Update step = 20' respectively represent the numbers of iterations in the inner loop as 1, 10, and 20.}
\label{fig12}
\end{figure*}

As we mentioned earlier, our algorithm during the meta-training phase involves an inner loop and an outer loop. In the inner loop, for each velocity distribution sampled from the support set, we update the network parameters through one or more steps of gradient descent, as shown in lines 5 to 8 of Algorithm 1. Hence, does the number of iterations in the inner loop affect our algorithm's performance? To answer this, we use the same training configuration and data as in Section 3.1 of the numerical examples and train two more meta-initialized parameters, with the inner loop executing 1 and 10 iterations, respectively. Similarly, we refine these two trained meta-initialized parameters on the overthrust model. Their loss curves compared to the Meta-PINN from Section 3.3 of the numerical examples are displayed in Figure \ref{fig12}. We observe that Meta-PINN 20 achieves the fastest convergence speed and highest accuracy, while Meta-PINN 1 exhibits faster convergence speed and accuracy than Meta-PINN 10. Therefore, the relationship between the number of iterations in the inner loop and the performance of Meta-PINN is not merely linear. In our practice, since using 20 iterations provide a superior performance, we use it as our initial parameter for the numerical examples section.

\subsection{Can we combine the Meta-PINN with other developments?}
We have demonstrated our method's advantages in convergence speed and accuracy on both layer model and the complex overthrust model. Here, we only utilize the simplest MLP architecture as our PINN baseline, without employing the existing advanced network architecture and training techniques. However, since our method constitutes an improvement in training methodology to improve generalization, it can be combined with any recent developments in network architecture, input embedding, loss functions, and so on. For example, Waheed et al. \cite{waheed2022kronecker} proposed that Kronecker neural networks with adaptive activation functions that enable PINNs to converge faster and provide more accurate wavefield simulations. Alkhalifah and Huang \cite{alkhalifah2023physics} suggested substituting the last hidden layer of the fully connected neural network model with a learnable Gabor functional solutions of the wave equation to enhance the efficiency and accuracy of neural network wavefield solutions. Furthermore, Huang and Alkhalifah \cite{huang2022pinnup} developed a training technique involving frequency upscaling and neuron splitting, and coupled it positional encoding (PE) applied to the network input, termed PINNup, to facilitate PINNs in providing more accurate wavefield solutions and faster convergence speeds. We can replace our MLP with these state-of-the-art network architecture  improvements to further enhance the performance of our method.

\begin{figure*}[!t]
\centering
\includegraphics[width=0.5\textwidth]{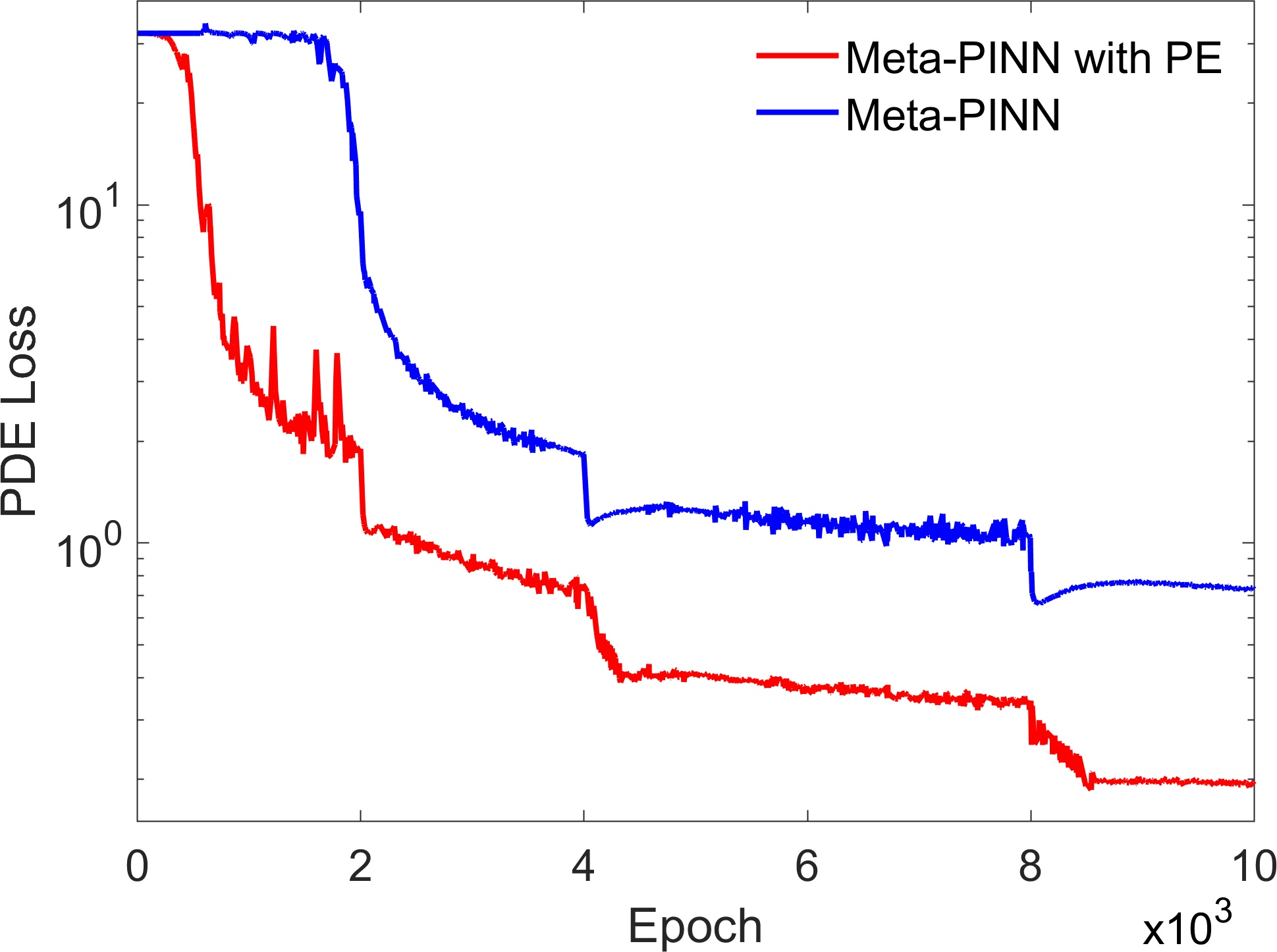}
\caption{A comparison of the physical loss curves between the Meta-PINN with positional encoding and without positional encoding.}
\label{fig13}
\end{figure*}

\begin{figure*}[!t]
\centering
\includegraphics[width=1\textwidth]{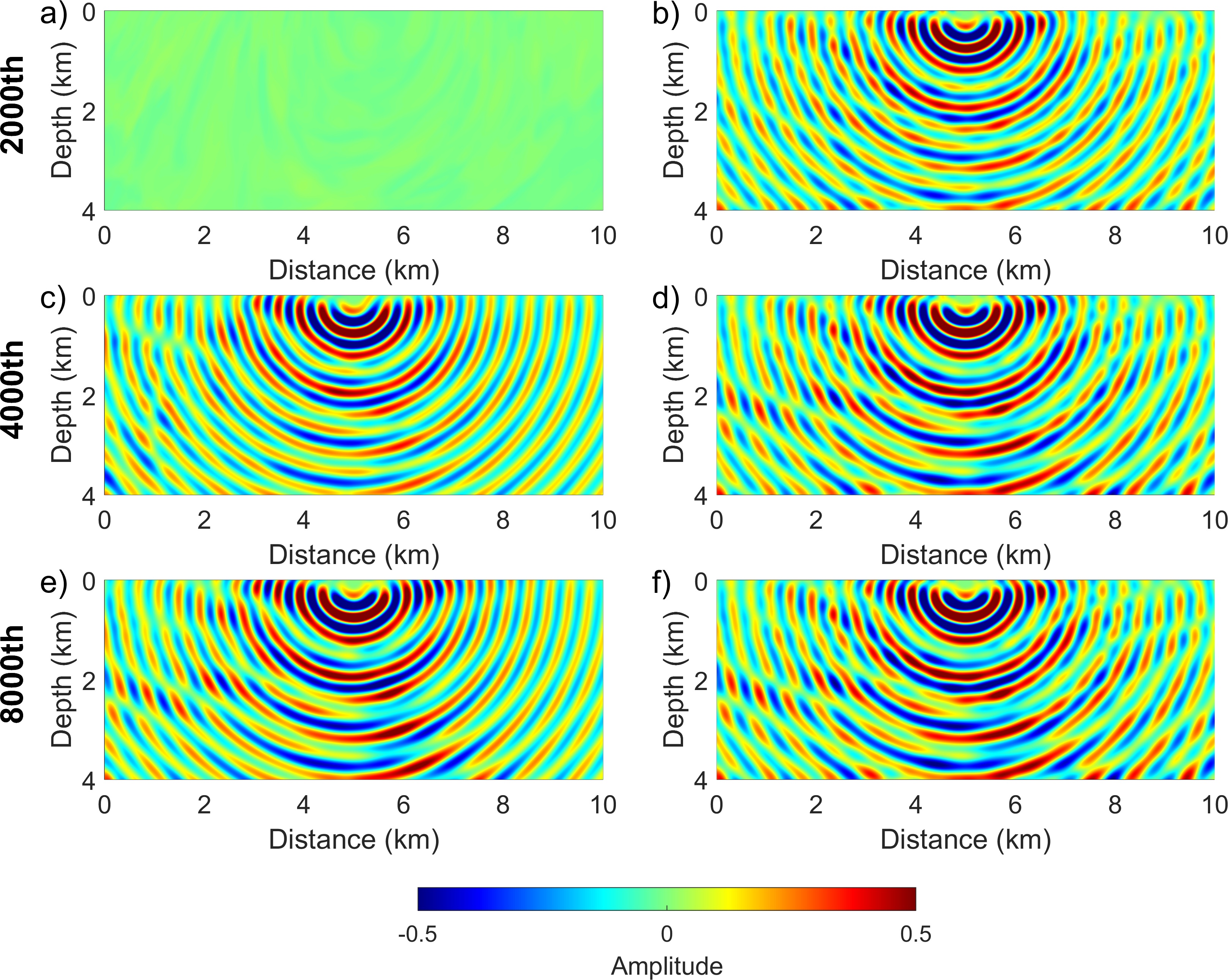}
\caption{The predicted real part of the scattered wavefield from the Meta-PINN with and without positional encoding after 2000 (fist row), 4000 (second row), and 8000 (third row) epochs of training for the overthrust model. The first and second columns correspond to the predicted scattered wavefield of Meta-PINN without and with positional encoding, respectively.}
\label{fig14}
\end{figure*}

To validate this, we make a slight modification in our network architecture by incorporating PE to test the performance with this architectural change. Due to the introduction of PE, we re-conduct the meta-training phase. Here, for the PE, we use $L=2$, which represents the dimensionality of the positional encoding vector for each input variable. The newly trained meta-initialized parameters also undergo the meta-testing phase on the overthrust model. The corresponding training loss curves are depicted in Figure \ref{fig13}, which is compared to the Meta-PINN without PE. We observe that with the addition of PE, our method's convergence speed has significantly improved, showing a rapid decrease in loss values from the outset, even for this complex overthrust model. Numerically, the Meta-PINN with PE reaches the accuracy of the Meta-PINN without PE trained for 10000 epochs in merely about 4000 epochs. We further compare the real part of the scattered wavefield solutions predicted by the Meta-PINNs with and without PE after 2000, 4000, and 8000 epochs of training, displayed in Figure \ref{fig14}. It is evident that with the aid of PE, the Meta-PINN could represent the main parts of the wavefield solution by the 2000th epoch, while the original Meta-PINN still could not provide a wavefield solution. With more extended updates, PE assists the Meta-PINN in delivering a more detailed wavefield solution. For instance, compared to the original Meta-PINN, the Meta-PINN with PE offers a more refined representation of the wavefield on the right side of the model, closer to the reference numerical wavefield. These results demonstrate that by integrating some existing methods, our Meta-PINN can exhibit superior wavefield modeling capabilities.

\subsection{What are the side effects?}
Certainly, every method has its limitations, and our approach is no exception. While our Meta-PINN framework demonstrates excellent performance in convergence speed and prediction accuracy, it relies on a preliminary meta-training phase to obtain robust initial network parameters. This meta-training stage involves a bidirectional gradient update, making it memory-intensive and time-consuming. Table \ref{tab1} displays the memory consumption and the time per epoch for meta-trainings with varying amounts of training data from Section 4.2. We can see that as the number of training samples increase, the memory and training time costs also rise. For instance, when training samples increase from 5 to 10, the memory usage and training time approximately double. If we are to increase the samples to 100, the memory usage would be around 580 GB, and the time per epoch would be 174 seconds. Although using only 10 velocity models through meta-learning provides a robust initialization with rapid adaptability, Section 4.2 suggests that incorporating more velocity models enables the Meta-PINN to converge faster and provide more accurate wavefield solutions. However, adding more training samples significantly increases memory usage and meta-training time. Therefore, future work needs to consider how to effectively reduce the memory and time burden of the meta-training phase without sacrificing performance.

\begin{table}
\centering
\caption{The memory and time consumption required for different amounts of training data during the meta-training phase.}
\renewcommand\arraystretch{1.5}
\setlength{\tabcolsep}{20pt}
\begin{tabular}{ccc}
    \hline
    \text {The amount of training data} & \text { Memory cost } & \text { Time cost } \\
    \hline
    \text {5} & $28.93$ \text {GB}  & $8.70$ \text {s} \\
    \text {8}  & $45.36$ \text {GB} & $13.92$ \text {s} \\
    \text {10} & $56.33$ \text {GB} & $17.40$ \text {s} \\
    \hline
\end{tabular}
\label{tab1}
\end{table}
\section{Conclusion}
Traditional numerical solvers, while stable, are often restricted by high computational costs and a fixed grid. In contrast, physics-informed neural networks (PINNs) offer grid-free and efficient solutions to complex wave equations but require extensive retraining for each new velocity model, leading to computational inefficiency. We proposed a Meta-PINN framework that addresses these challenges by leveraging a meta-learned network initialization, facilitating rapid adaptation and significantly reducing the computational resources needed for modeling different velocity distributions. The results from our numerical examples validate the efficacy of the Meta-PINN approach. Through a comprehensive training and testing process involving various velocity models, Meta-PINN demonstrated not only accelerated convergence in training but also heightened accuracy in wavefield solutions when compared to vanilla PINN. These findings underscore the potential of Meta-PINN as an advanced tool for seismology, offering a significant step forward in efficient and accurate seismic modeling.
\section*{Acknowledgments}
This publication is based on work supported by the King Abdullah University of Science and Technology (KAUST). The authors thank the DeepWave sponsors for supporting this research. This work utilized the resources of the Supercomputing Laboratory at King Abdullah University of Science and Technology (KAUST) in Thuwal, Saudi Arabia.
\section*{Code and Data Availability}
The accompanying codes that support the findings will be shared upon acceptance of this paper.

%Bibliography
\bibliographystyle{unsrt}  
\bibliography{references}

\end{document}